# Discrete, recurrent, and scalable patterns in human judgement underlie affective picture ratings

Emanuel A. Azcona[1,†], Byoung-Woo Kim[2,†], Nicole L. Vike[2,§], Sumra Bari[2,§], Shamal Lalvani[1,§], Leandros Stefanopoulos[1,§], Sean Woodward[2,§], and Martin Block[3,§], Aggelos K. Katsaggelos[1,4,5,‡], Hans C. Breiter[2,6,‡]

[1] Image and Video Processing Lab, Department of Electrical and Computer Engineering, Northwestern University, Evanston, IL, USA
[2] Department of Psychiatry and Behavioral Sciences, Northwestern University, Chicago, IL, USA
[3] Integrated Marketing Communications, Northwestern University, Evanston, IL, USA
[4] Department of Computer Science, Northwestern University, Evanston, IL, USA
[5] Department of Radiology, Northwestern University, Chicago, IL, USA
[6] Laboratory of Neuroimaging and Genetics, Department of Psychiatry, Massachusetts General Hospital and Harvard School of Medicine, Boston, MA, USA

† These authors share first authorship
§ These authors share second authorship
‡ These authors share corresponding/senior authorship

**\*Correspondence:**

For Feature Engineering, Project Design and Management: Hans C. Breiter, Laboratory of Neuroimaging and Genetics, MGH Division of Psychiatric Neuroscience, 120 2nd Avenue, Charlestown, MA 02129-2060, USA. Email: hbreiter@mgh.harvard.edu

For Statistical Analysis: Aggelos K. Katsaggelos, Department of Electrical and Computer Engineering, 2145 Sheridan Road, Tech Room M468, Evanston, IL 60208, USA. Email: a-katsaggelos@northwestern.edu

## Abstract

Operant keypress tasks, where each action has a consequence, have been analogized to the construct of "wanting" and produce lawful relationships in humans that quantify preferences for approach and avoidance behavior. It is unknown if rating tasks without an operant framework, which can be analogized to "liking", show similar lawful relationships. We studied three independent cohorts of participants ($N = 501$, $506$, and $4{,}019$ participants) collected by two distinct organizations, using the same 7-point Likert scale to rate negative to positive preferences for pictures from the International Affective Picture Set. Picture ratings without an operant framework produced similar value functions, limit functions, and trade-off functions to those reported in the literature for operant keypress tasks, all with $R^2$ goodness of fits above 0.75. These value, limit, and trade-off functions were discrete in their mathematical formulation, recurrent across all three independent cohorts, and demonstrated scaling between individual and group curves. In all three experiments, the computation of loss aversion showed 95% confidence intervals below the value of 2, arguing against a strong overweighting of losses relative to gains, as has previously been reported for keypress tasks or games of chance with calibrated uncertainty. Graphed features from the three cohorts were similar and argue that preference assessments meet three of four criteria for lawfulness, providing a simple, short, and low-cost method for the quantitative assessment of preference without forced choice decisions, games of chance, or operant keypressing. This approach can easily be implemented on any digital device with a screen (e.g., cellphones).

# 1   Introduction

Preference can be defined as the variable extent an organism shows an inclination to act or behave by approaching or avoiding events in the world, based on the rewarding or aversive effects of these events (Lewin, 1935; Schneirla, 1959, 1965). Preference-based behavioral variables that measure the intensity and patterns of approach/avoidance behavior with an operant keypress task show lawful relationships in humans when using visual (Kim et al., 2010; Lee et al., 2015) and auditory stimuli (Livengood et al., 2017) (see Fig. 1). These lawful behavioral relationships (Feynman, 1965) have been associated with activation in brain reward circuitry by use of model-based functional MRI (Aharon et al., 2001; Viswanathan et al., 2015), imaging genetics (Gasic et al., 2009; Perlis et al., 2008), and quantitative morphometry (Makris et al., 2008).

The keypress task used in such studies was derived from an operant framework where each keypress had an incremental consequence on stimulus view time (Aharon et al., 2001; Lee et al., 2015); this has been well-validated across multiple studies (Aharon et al., 2001; Elman et al., 2005; Gasic et al., 2009; Kim et al., 2010; Lee et al., 2015; Levy et al., 2008; Makris et al., 2008; Perlis et al., 2008; Strauss et al., 2005; Viswanathan et al., 2015, 2017; Yamamoto et al., 2009). The keypress task can be analogized to the construct of "wanting" as opposed to "liking" (Aharon et al., 2001; Berridge & Robinson, 2016), and leads to variables that quantify the average (mean) magnitude ($K$), variance ($\sigma$), and the pattern (i.e., Shannon entropy ($H$)) of participants' keypress-based behavior. We refer to this methodology, and the multiple relationships between these variables and features based on their graphical relationships, as relative preference theory (RPT) (Fig. 1). Two features of RPT mimic known functions with distinct variables from prospect theory (Kahneman & Tversky, 1979) and the mean-variance function described by Markowitz (1952) for portfolio theory.

To date, RPT has only been discussed in an operant framework where effort traded for viewing time can be considered a model of "wanting" (e.g., Aharon et al., 2001). In a recent study, RPT was compared to a prospect theory framework in which ratings were made under conditions of uncertainty during a game of chance (i.e., anticipation phase of the trial), and under conditions of certainty when the outcome was revealed (i.e., outcome phase of the trial). During anticipation, ratings produced statistically similar loss aversion (*LA*) measures to those of keypressing with an RPT analysis, whereas during the outcome phase, ratings showed no overweighting of losses relative to gains (Lee et al., 2015). *LA* was specifically defined by Tversky and Kahneman (1992) to describe an overweighting of negative judgements relative to positive ones under conditions of risk. These observations raised the hypothesis that in a non-operant model where actions have no consequences (i.e., "liking"), a rating task with no uncertainty might produce RPT-like curves, but not show the same degree of overweighting of losses relative to gains which characterize *LA* during uncertainty in prospect theory. Demonstrating that rating tasks show consistent law-like patterns, but a reduction in the overweighting of negative outcomes for *LA*, has potential implications for online digital behavior. Specifically, an absence of strong *LA* in the context of liking responses, but presence with wanting responses, provides a potential hypothesis for why digital behavior that is not effort or operant based, might reflect less concern for negative consequences.

These considerations led us to analyze three separate cohorts of human participants using a rating task and picture stimuli from the International Affective Picture Set (IAPS) (Lang et al., 2008). Three questions were asked:

(1) Would ratings in the absence of the operant framework, produce preference relationships similar to what has been observed with keypressing (e.g., Fig. 1)?

(2) If ratings produced RPT-like curves, would these functions be (i) mathematically discrete, (ii) recurrent across cohorts and the number of pictures studied, and (iii) scale from individual to group data? Namely, would graphs from ratings meet three of the primary criteria raised by Feynman (1965) for lawfulness?

(3) How would these curves differ from previously published functions observed with keypressing, and would features of these curves, like those shown in Fig. 1, be consistent across cohorts?

The broad set of features extracted from rating tasks in this study provide a potential framework for summarizing human approach and avoidance tendencies in judgement decisions. This set of summary metrics, derived from a simple rating task on a digital device, has the potential to characterize human preference at the big-data scale, with thousands of participants.

## 2   Methods

### 2.A   Participants

In all three studies, rating and survey responses were collected online to meet demographic criteria established by the U.S. census. Three studies were performed. One study involved 501 participants for the Emotion and Behavior Study (EBS), an online study of  United States (U.S.) adult (i.e., ≥ 18 years of age) consumers, conducted by Research Results, Inc. (Boston, MA) in 2016. The second study consisted of 506 participants randomly sampled from the general U.S. population using a participant database accessed by Gold Research Inc. (San Antonio, Texas) for the Automated Mental Health Assessment Study (AMHA), referred to as the AMHA-1 cohort. Questionnaire responses for AMHA-1 were collected between the end of February 2021 and the beginning of March 2021, approximately one year following the official pandemic declaration in the U.S. The third study involved 4,019 participants, also randomly sampled from the general U.S. population using a participant database accessed by Gold Research Inc. for the AMHA study from November 2021, referred to as the AMHA-2 cohort. All participants provided informed consent for their response data to be used, and data released to Northwestern University from Gold Research Inc. and Research Results Inc. were anonymized.

Participant demographic information including gender identity, age group (in years), household income, employment status, education level, handedness, and race/ethnicity are summarized in Tables 1a-c, including percent compositions for each group within each corresponding demographic measure.

## 2.B  Picture Stimuli

Stimulus sets across the rating task consisted of images from the International Affective Picture System (IAPS) (Lang et al., 2008), a well-validated emotional stimulus set. For all cohorts, six categories of pictures were used: (1) sports, (2) disasters, (3) cute animals, (4) aggressive animals, (5) nature (beach vs. mountains), and (6) food, with eight pictures per category (48 pictures in total). Pictures had a maximum size of $1{,}204 \times 768$ pixels in all three studies. All picture stimulus sets reported in the present study are collectively referred to as "IAPS stimuli" throughout the text.

## 2.C  Picture Rating Task ("Liking" Assessment)

Participants were prompted for the rating task while completing an online digital survey, which contained questionnaires regarding participant demographic information and research questionnaires for depression symptoms using the Patient Health Questionnaire (PHQ-9) (Kroenke et al., 2001); trait anxiety using the Spielberger State-Trait Anxiety Inventory (STAI) (Spielberger et al., 2001); a broad array of mental health, neurological, and medical issues using the MGH Phenotype Genotype Project in Addiction and Mood Disorders symptom questionnaire (MGH-SQ); and behavioral health disorders (e.g., internalizing or externalizing psychiatric disorders, substance use disorders, or crime/violence problems) from the GAIN-SS short screen assessment (Dennis et al., 2006). For the picture rating task, the instructions presented to participants were as follows:

> *"The next part of this survey involves looking at pictures and then responding how much you like or dislike the image. Please rate each image on a scale from -3 (Dislike Very Much) to +3 (Like Very Much). Zero (0) is neutral... meaning you have no feelings either way. The images are a set of photographs that have been used by scientists around the world for over 20 years.*
>
> *It is important you rate each picture based on your initial emotional response.*
>
> *There are no right or wrong answers... just respond with your feelings and rate the pictures very quickly.*

*Please click "Next" to begin."*

Each picture was presented as shown in Fig. 2, where the ratings below each picture were selectable using the mouse cursor. There was no time limit for assigning ratings to each picture, but participants were requested to rate each picture as quickly as possible, and they were not able to change their response after selecting a rating. After each rating selection was made, the next picture was automatically loaded and presented.

## 2.D    Data Quality Screening

Data integrity was assessed for all data from the three studies. Quality assurance was conducted based on four exclusion criteria for picture rating tasks and survey data (non-rating data not described herein), which reduced the analysis to 281 participants for the EBS cohort, 366 for AMHA-1, and 3,476 for AMHA-2. These four exclusion criteria were:

(1) Participants selected the same response throughout any section of the questions/tasks (e.g., selecting option "1" for all questions),

(2) Participants indicated they had ten or more clinician-diagnosed illnesses (data not described here),

(3) Participants showed minimal variance in a picture rating task (i.e., all pictures were rated the same or varied only by one point; data not described here), and

(4) If *both* education level and years of education did not match *and* if they completed the questionnaire in less than 500 seconds (800 seconds for the AMHA-2 cohort).

Further quality assurance involved assessment of RPT variables and curves from the picture rating tasks. Variables that were quantified included the average magnitude ($K$), variance ($\sigma$), and the pattern or information (i.e., Shannon entropy ($H$)) related to participants' preference behavior. $K$ reflected the average (mean) rating a subject made to either approach ($K_+$) or avoid ($K_-$) stimuli within each picture category. Other metrics included the variance to approach ($\sigma_+$) or avoid ($\sigma_-$) stimuli, along with the Shannon entropy (i.e., information; see Shannon & Weaver, 1949) of ratings describing approach ($H_+$) or avoidance ($H_-$) of stimuli within each category. The Shannon entropy is a core variable in information theory that characterizes the degree of uncertainty across a set of responses (Shannon & Weaver, 1949); it quantifies the pattern of judgements made to a set of stimuli and could thus be considered a memory variable. Collectively, these variables capture judgments about the valence of judgement (approach or avoidance) as well as its magnitude (intensity of rating) to describe relative preferences (Kim et al., 2010) (Fig. 1).

When evaluating data quality, raw data was assessed for cases when $K = 0$ for a given category (i.e., cases where the subject made all neutral ratings to neither approach nor to avoid any stimulus in the category). Computing the Shannon entropy, $H$, for a given picture category requires that $K > 0$ given that when $K = 0$, the $H$ computation results in evaluating $\log\left(\frac{0}{0}\right)$, which is undefined. In these cases, the Shannon entropy was set to $H = 0$ for categories in which the subject rated "0" for all the stimuli.

Before carrying out the RPT analyses, and fitting models to participants' ratings, data was further screened for additional criteria beyond when $K = 0$ for a given category. The complete set of model fit inclusion/exclusion criteria was as follows:

(1) Valid entropy ($H$) calculations (see prior paragraph),

(2) Exclusion of data points lying beyond three times the interquartile range (IQR), below the first quartile or above the third quartile (i.e., removing extreme outliers),

(3) A sufficient number of data points to fit the model with a computable $R^2$ (e.g., at least three points for a non-linear fit), and

(4) Coherence of model fits between individual and group data. This last criterion required that the curve direction for individual subject fits be consistent with the curve direction of the group-level statistical fits (and boundary envelopes), and therefore corroborate most of the observed subject data.

Criteria (3) and (4) are necessary operational definitions for quality assurance given the potential for convergence failures with curve fitting. For the AMHA-2 cohort, criteria (3) was ignored since the exclusion of three times the IQR would have resulted in too much data loss.

In total, six types of model fitting were performed for the rating data: group and individual models for the $(K, H)$ data, $(K, \sigma)$ data, and $(H_+, H_-)$ data distributions. For the group data, we generated group-level data fits along with boundary envelopes (power-law fits and logarithmic fits for group $(K, H)$ data), and quadratic fits for group $(K, \sigma)$ data to guide the focus of statistical testing based on the power law fits $(K, H)$, and quadratic fits $(K, \sigma)$ for individual data. Individual data then followed this based on logarithmic and simple power-law fits for individual $(K, H)$ value functions, quadratic fits for individual $(K, \sigma)$ limit functions, and radial fits for individual $(H_+, H_-)$ tradeoff distributions (Kim et al., 2010; Livengood et al., 2017).

## 2.E   Relative Preference Analysis

For the relative preference analysis, we replicated the methodology described in detail by Kim et al. (2010), Livengood et al. (2017), and Viswanathan et al. (2017). We used the iterative modeling approach of Banks and Tran (2009) to identify RPT patterns in the data and three of the four putative signatures of lawfulness, as described previously with visual (Breiter & Kim, 2008; Kim et al., 2010) and auditory stimuli (Livengood et al., 2017). We thus sought "discrete" mathematical fitting of patterns within the data, "recurrence" of patterns across the three distinct experimental cohorts, and "scalability" of the observed patterns. We utilized datasets that met stringent criteria for quality assurance, then assessed the graphical structure between the three behavioral variables $\{K, H, \sigma\}$. For the rating tasks, these variables reflected the mean positive ratings or negative ratings within a picture category $(K_+, K_-)$, the Shannon entropy of positive/negative ratings within a category $(H_+, H_-)$, and the standard deviation $(\sigma_+, \sigma_-)$ of positive/negative ratings within a category. Graphical analyses sought to determine the presence of functions, manifolds, or boundary envelopes to individual, and separately, group data that were graphically similar to RPT functions, manifolds, and boundary envelopes (Breiter & Kim, 2008; Kim et al., 2010; Lee et al., 2015; Livengood et al., 2017).

Formal testing of discreteness, recurrence, and scaling was done as follows. To assess if mathematical fitting was discrete, the goodness of fit for the $(K, H)$ value functions and $(K, \sigma)$ limit functions, across the three experiments, were characterized by $R^2$, adjusted $R^2$, and $F$-test statistics; then tabulated by location and dispersion estimates. Given prior keypress findings of discreteness with $R^2 > 0.7$, we assessed if definable functions for individual data and manifold fits (and/or boundary envelopes) for group data had clear parameter estimates and showed $R^2 > 0.7$. For recurrence, we assessed if similar individual and group models were observed for each of the three independent populations, and if the extracted RPT features ($N = 15$) for individual functions were similar across the three groups. Lastly, scale invariance and simple power-law fitting was assessed by performing linear regressions following logarithmic transformations of both the $K$- and $H$-axes. If the resulting fits characteristically demonstrated asymptotic behavior ($0 < a < 1$, given $H(K) = bK^a$), this implied that substantial changes in the input variable, $K$, produced only minor changes in the output, $H$. The same asymptotic behavior was assessed with the logarithmic fits to the $(K, H)$ data, with the difference that in this case the fits were obtained by performing a linear regression of $H$ against $K$ after the logarithmic transformation of $K$ alone.

## 2.E.1   $(K, H)$ value functions

We evaluated mean positive or negative ratings across stimuli within a picture category $(K_+, K_-)$ and the Shannon entropy of these ratings $(H_+, H_-)$. We used the following approach to compute the Shannon entropy separately for the positive (approach) and negative (avoidance) ratings in each category. First, consider an ensemble of numbers for either positive or negative rating responses, $\mathbf{a}$, across stimuli within a single picture category: $\mathbf{a}_\pm = \{a_1, a_2, \dots, a_N\}$, for $N$ pictures within the given category. We can then define the relative proportions of the positive and negative responses for the individual stimuli, $p_i$, such that

$$p_i = \frac{a_i}{\sum_{j=1}^{N} a_j}. \tag{1}$$

Using these normalized proportions of the rating responses, the Shannon entropy of the response pattern can be computed for an individual picture category as follows:

$$H_{\pm} = \sum_i p_i \log_2 \left(\frac{1}{p_i}\right). \tag{2}$$

After computing the values of $K_{\pm}$ and $H_{\pm}$ for each picture category, $(K, H)$ value functions were generated by plotting the Shannon entropy, $H_{\pm}$, against the mean ratings ($K_{\pm}$), for all picture categories for an individual subject. $(K, H)$ data were also plotted across multiple participants to visualize data at the group level.

At the group level, we assessed if $(K, H)$ best-fit parameters could be approximated using the logarithmic function, $H(K) = a \log_{10}(K) + b$, or power law functions, $H(K) = bK^a$; we also confirmed they contained boundary envelopes that conformed well to either logarithmic functions of power-law functions. At the individual subject level, we assessed fits for the same logarithmic and power-law functions to the $(K, H)$ data for approach and avoidance across picture categories for individual participants. The best-fit parameters for the logarithmic and power-law functions were achieved by performing a simple linear regression on the plots for $H$ vs. $\log_{10}(K)$, and $\log_{10}(H)$ vs. $\log_{10}(K)$, respectively.

### 2.E.2 $(K, H)$ limit functions

The second relationship considered was that between the mean ratings, $K_{\pm}$, and the standard deviation of ratings across stimuli within a category, $\sigma_{\pm}$. $(K, \sigma)$ limit functions were generated by plotting values of $\sigma$ against $K$ for all picture categories in an individual subject or by pooling the data together across participants in a group analysis. At both the individual and group level, we found that $(K, \sigma)$ limit functions were well characterized by quadratic functions of the form $\sigma = aK^2 + bK + c$. For the group data, we fit quadratic boundary envelopes to the $(K, \sigma)$ data much in the same manner performed for the $(K, H)$ value functions. For individual subject analysis, we fit quadratic functions directly to the $(K, \sigma)$ data using the polyfit() function in MATLAB®.

### 2.E.3 $(H_+, H_-)$ trade-off plots

$(H_+, H_-)$ trade-off (or opponency) plots were defined by plotting the Shannon entropy for positive ratings, $H_+$, against the Shannon entropy for negative ratings, $H_-$, for all picture categories in each stimulus set. These plots were generated either across categories for an individual subject, or by pooling data across all participants in the cohort to generate a group-level plot. For both the individual subject- and group-level data, $(H_+, H_-)$ data conformed to a radial distribution about the origin of the trade-off plot, such that $r = \sqrt{(H_+)^2 + (H_-)^2}$, or equivalently, $H_+ = \sqrt{r^2 - (H_-)^2}$. Radial fits were estimated for individual participants as well as the group-level data by computing the mean radial distance, $r$, across all $(H_+, H_-)$ data in the trade-off plot.

### 2.E.4 Feature extractions from $(K, H)$, $(K, \sigma)$, and $(H_-, H_+)$ functions

To help characterize the $(K, H)$, $(K, \sigma)$, and $(H_-, H_+)$ functions, two standard definitions from behavioral economics and 13 other features that reflect standard curve feature analyses were utilized (Fig. 1). The two behavioral economic measures were loss and risk aversion, $LA$ and $RA$ respectively. In parallel with these, we defined 13 other curve features: loss resilience ($LR$), positive offset ($\beta_+$), negative offset ($\beta_-$), positive apex ($\alpha_+$), negative apex ($\alpha_-$), positive turning point ($\rho_+$), negative turning point ($\rho_-$), mean polar angle of the $(H_-, H_+)$ curve ($\theta$), standard deviation of the polar angle ($\sigma_\theta$), mean radial distance of points on the $(H_-, H_+)$ plot ($r$), and the standard deviation of the radial distances ($\sigma_r$). Descriptions of these 15 curve features are described briefly in what follows.

### 2.E.4.1 $(K, H)$ extracted feature definitions

Features for the $(K, H)$ plots, were framed by the $(K, H)$ function being considered as a power-law function, that was concave relative to the $K$-axis. The RPT features that were extracted from these graphs were: risk aversion ($RA$), loss resilience ($LR$), loss aversion ($LA$), and the positive and negative offsets ($\beta_+$ and $\beta_-$, respectively).

- *Risk aversion (RA)*: risk aversion is extracted as the ratio of the second derivative of the $(K_+, H_+)$ curve to its first derivative, which also produces a curve. To produce a unitary value for comparison across cohorts, we calculated $RA$ for $K_+ = 0.5$. Informally, $RA$ measures the degree to which an individual prefers a likely reward in comparison to a better more uncertain reward. $RA$ is a common notion in economics that studies decision-making under uncertainty (Zhang et al., 2014).

- *Loss resilience (LR)*: loss resilience is defined to be the absolute value of the ratio of the second derivative of the $(K_-, H_-)$ curve to its first derivative, which also produces a curve. For prediction, we calculated $LR$ at $K_- = 1.5$. Informally, $LR$ is the degree to which an individual prefers to lose a small defined amount in comparison to losing a greater amount with more uncertainty associated with this loss.

- *Loss aversion (LA)*: loss aversion is the absolute value of the ratio of the linear regression slope of $(\log K_-, \log H_-)$ to the linear regression slope of $(\log K_+, \log H_+)$. It intuitively measures the degree to which an individual person overweighs losses to gains. $LA$ a fundamental measure in Kahneman and Tversky's (1979) prospect theory, which informally states that humans have a cognitive bias to overweight losses relative to gains in the presence of uncertainty.

- *Positive offset ($\beta_+$)*: the positive offset is the value of $K_+$ when setting $H_+ = 0$. $\beta_+$ intuitively measures the ante one needs to engage in a game of chance and models the amount of a bid an individual is willing to make to enter a game of chance (e.g., poker).

- *Negative offset ($\beta_-$)*: the negative offset is the value of $K_-$ when setting $H_- = 0$. $\beta_-$ intuitively measures how much insurance an individual might need against bad outcomes. It mirrors the ante, but in the framework of potential losses.

### 2.E.4.2   $(K, \sigma)$ extracted features definitions

    Features for the $(K, \sigma)$ curves, were framed by the $(K, \sigma)$ curves being considered as quadratic functions, that were concave relative to the $K$-axis. The $(K, \sigma)$ curve models the relationship between variance (risk) and mean value. It can also be framed by the following question: Would an individual prefer a dollar with probability one, or value drawn from a normal distribution with a mean of two and variance of two? The RPT features that are extracted from this curve include: the positive and negative apices ($\alpha_+$ and $\alpha_-$, respectively), the positive and negative turning points ($\rho_+$ and $\rho_-$, respectively), and the positive and negative quadratic areas ($QA_+$ and $QA_-$, respectively).

- *Positive apex ($\alpha_+$)*: the positive apex is the value of $\sigma_+$ for the derivative $\frac{d\sigma_+}{dK_+} = 0$. Intuitively, this represents the maximum variance for approach behavior. In this sense, $\alpha_+$ models where increases in positive value transition from a relationship with increases in risk, to a relationship with decreases in risk. Markowitz (1952) described decision utility similarly, so that the positive apex models when variance changes from weighing against a decision to facilitating a decision.

- *Negative apex ($\alpha_-$)*: the negative apex is the value of $\sigma_-$ when the derivative $\frac{d\sigma_-}{dK_-} = 0$. Intuitively, this represents the maximum variance for avoidance behavior. Like with the positive apex, this transition point is important to consider for avoidance decisions in the context of Markowitz's (1952) decision utility.

- *Positive turning point ($\rho_+$)*: the positive turning point is the value of $K_+$ when the derivative $\frac{d\sigma_+}{dK_+} = 0$. Intuitively, this represents the rating intensity with maximum variance for approach behavior, potentially when an individual decides to approach a goal-object.

- *Negative turning point ($\rho_-$)*: the negative turning point is the value of $K_-$ when the derivative $\frac{d\sigma_-}{dK_-} = 0$. Intuitively, this represents the rating intensity with maximum variance for avoidance behavior, potentially when an individual decides to avoid a goal-object.

- *Positive quadratic area ($QA_+$)*: the positive quadratic area is the area under the curve (AUC) of the first quadrant of the $(K_+, \sigma_+)$. This variable represents the relationship between $K_+$ and $\sigma_+$ and can be thought of as a quantity that measures the amount of value an individual associates to positive stimuli.

- *Negative quadratic area ($QA_-$)*: the negative quadratic area is the AUC of the third quadrant for the curve $(K_-, \sigma_-)$. This variable represents the relationship between $K_-$ and $\sigma_-$ and can be thought of as quantity that measures the aversive value an individual associates to negative stimuli.

### 2.E.4.3 <u>$(H_-, H_+)$ extracted features definitions</u>

Features for the $(H_-, H_+)$ curve were framed as the $(H_-, H_+)$ function being considered a trade-off function between the $H_+$ and $H_-$ variables, that can commonly look like a semi-circular fit in individuals (e.g., Fig. 1c). The RPT features = extracted from this curve include: the mean polar angle ($\theta$), its standard deviation ($\sigma_\theta$), mean radial distance ($r$), and its corresponding standard deviation ($\sigma_r$).

- *Mean polar angle ($\theta$)*: the mean polar angle is the mean of the polar angles of the points in the $(H_-, H_+)$ plane. Intuitively, this measures the mean balance entropies, or patterns, in approach to avoidance behavior.

- *Polar angle standard deviation ($\sigma_\theta$)*: the standard deviation of the polar angles of the points in the $(H_-, H_+)$ plane. Intuitively, this measures the standard deviation in the patterns of approach to avoidance behavior. This variance represents the spread of positive and negative preferences across a set of potential goal-objects and can be considered a measure of the breadth of an individual's (or a group's) preferences.

- *Mean radial distance ($r$)*: the mean radial distance measures the average Euclidean distance of the data points in the $(H_-, H_+)$ curve to the origin. This measure defines how individuals can have strong preferences (i.e., biases) for the same thing, reflecting *conflict*, or having low preferences for something, reflecting *indifference*. This gets at the consistency of compatibility of approach and avoidance, and how an individual can both like and dislike something or be indifferent to both its positive and negative features.

- *Radial distance standard deviation ($\sigma_r$):* this is simply the standard deviation of the radial distances of the data points in the $(H_-, H_+)$ plane to the origin. This final measure, is interpreted through how the points in the $(H_-, H_+)$ plane vary regarding the radial distance from the origin. The variance in this radial distance will reflect how much an individual goes between having *conflicting* preferences and having *indifferent* ones.

## 2.F  *Comparison of Features between Rating Experiments*

For each of the three subject populations, the mean and standard deviation (SD) were computed for each of the fifteen features, along with standard error of the mean (SEM) and the 95% confidence intervals (CI) for the corresponding means. Violin plots (Hintze & Nelson, 1998) for each of the RPT features were also generated (Fig. 7) to provide a visual comparison of the distribution, interquartile range (IQR), and 95% CIs, with respect to the corresponding median, for each RPT feature across all cohorts. The primary framework of comparison was assessment of overlap in the 95% CI for the corresponding means, and violin plots for the medians.

A quantitative comparison of RPT features across the three cohorts was also performed using rank-based, nonparametric Kruskal-Wallis *H*-test (Kendall & George, 2008b) statistics, followed by *post-hoc* nonparametric pairwise multiple comparisons using Dunn's test (Dinno, 2015) and Kolmogorov-Smirnov test (Kendall & George, 2008a) statistics. This was done for all fifteen features, although only seven of these features reflected dimensionless units, and caveats around demographic differences in the cohorts could not be incorporated into such analyses.

# 3 Results

## 3.A Group-level $(K, H)$, $(K, \sigma)$, and $(H_-, H_+)$ Analyses

We first investigated the relationships between mean ratings and the Shannon entropy of category distributions for ratings. Group-level analyses were performed in two ways:

(1) Envelope fits as done previously (e.g., Kim et al., 2010; Livengood et al., 2017), and

*(2)* statistical fits of group data to constrain the functional fits test subsequently with individual data analyses (Table 3a-c).

For envelope fitting of the value function, power-law and logarithmic boundary envelopes were fit to the approach $(K_+, H_+)$ and avoidance $(K_-, H_-)$ rating data such that they formed an outer bound containing 95% of the data; they were both observed to provide robust approximations of the edge of the distribution. For all three experiments, group data was fit by boundary envelopes to similar extents (all $p < 0.05$) by both logarithmic and power-law functions. When we examined functional fitting of the group data, we observed statistically significant fits (all $p < 0.05$) by both logarithmic and power-law functions (Tables 3a-c).

Next, we examined the relationship between the average category ratings and the standard deviation of ratings in each category, which we refer to as the mean-variance relationship (Fig. 1b). Boundary envelopes enclosing 95% of the approach $(K_+, \sigma_+)$ and avoidance $(K_-, \sigma_-)$ data were fit to the EBS data. The quadratic boundary envelopes effectively approximated the edge of the mean-variance plots. The same analytic approach was performed for AMHA-1 and AMHA-2 data, showing outer bounds containing 95% of the data, providing robust approximations of the edge of the distribution, and broadly corroborating the behavior of analogous distributions for reported keypress-based RPT variables (Kim et al., 2010; Livengood et al., 2017), with $p < 0.05$ for all three datasets. When we examined functional fitting of the group data, we observed statistically significant fits (all $p < 0.05$) for all three cohorts (Tables 3a-c). Importantly, all curves depicting limit functions with group data showed concave fits (relative to the absolute vale of the ($K$, $H$, and $\sigma$ variables)), thereby setting a constraint used for the individual data.

Lastly, we examined the ($H_-$, $H_+$) tradeoff distributions characterizing the relationship between the patterns of approach and avoidance across tasks. This analysis sought to assess whether the pattern of approach preference behavior (i.e., positive ratings) scaled in proportion to the avoidance preference behavior (i.e., negative ratings) for pictures within the same categories. Specifically, we fit radial functions to test for symmetry in the distribution of $H_-$ and $H_+$ values across categories within each individual subject. Ratings-based group-level ($H_-$, $H_+$) distributions were broader than distributions we have reported previously for keypress-based implementations of the IAPS relative preference task, consistent with the increased variance in the rating graph features for mean polar angle and mean radial distance.

## 3.B  *Individual Subject* $(K, H)$ *Value Functions*

After investigating the shape of distributions at the group level, $(K, H)$ data and value function fitting at the individual subject level were assessed. As observed for group data, individual participants' $(K, H)$ value functions for the rating were well fit by concave logarithmic or power-law functions (Table 4, Figs. 3-5). Goodness of fit was assessed by computing $R^2$ values, adjusted $R^2$ values (accounting for degrees of freedom), and $F$-statistics for each subject's model fit (Table 4). $R^2$ values were all $> 0.75$ and extended up to 0.94. Average goodness of fit values were quite similar between the three cohorts (Table 4).

Overall, fewer participant exclusions were noted across the three studies for logarithmic fitting of the $(K, H)$ approach data when compared to $(K, H)$ power-law model fits due to either insufficient data for fitting of joint RPT distributions or invalid parameter estimates (see   Methods).

### 3.C   Individual Subject $(K, \sigma)$ Limit Functions

The consistency of mean-variance relationships across participants in the rating experiment was assessed by fitting quadratic functions to each individual subject's $(K, \sigma)$ distributions for approach and avoidance rating data. Concave quadratic fits across individual participants' $(K, \sigma)$ data are displayed in Figs. 3-5. As with the $(K, H)$ data, the goodness of fit was assessed by computing $R^2$ values, adjusted $R^2$ values (accounting for degrees of freedom), and $F$-statistics for each subject's model fit (Table 4). Adjusted $R^2$ values varied from 0.88 to 0.90, which was considered very high across participants.

### 3.D   Individual Subject $(H_-, H_+)$ Trade-off distributions

Trade-off distributions $(H_-, H_+)$ for individual participants' rating patterns across pictures were also examined. This analysis sought to assess whether the pattern of participants' approach preference behavior (i.e., positive ratings) scaled in proportion to the pattern of participants' avoidance preference behavior (i.e., negative ratings) for pictures within the same categories (e.g., nature scenes). Specifically, we fit radial functions to test for trade-offs in the distribution of $H_-$ and $H_+$ values across categories within each individual subject. Figs. 3-5 display radial fits across individual participants' $(H_-, H_+)$ data and highlight the $(H_-, H_+)$ data points and fit for a representative subject from each experiment.

### 3.E   Extracted Curve Features Computation and Comparison of Rating Results

Summary statistics for $LA$, $RA$, and the 13 other RPT graph features obtained for each subject in each experiment are summarized in Table 5; these features were statistically evaluated and are summarized in Tables 6a & 6b. Given Fig. 7, where the distributions of each RPT feature are displayed, we observe that the majority of the RPT features' distributions showed significant deviation from normality, nonparametric statistics were used and tabulated in Tables 6a & 6b. Most importantly, Fig. 7 demonstrates consistent overlap among the 95% CIs for the median for a majority of the RPT features.

$LA$ values were lower in all three cohorts than has been reported for keypress experiments and prospect theory-based experiments from Tom et al., 2007 and Tversky & Kahneman, 1992. It should be noted that $LA < 2.0$ for the rating experiments suggests that participants do not show $LA$ per se, but potential reward sensitivity. $LA$ values did differ between the three cohorts by the Kruskal-Wallis $H$-test, while the post-hoc Dunn's assessment confirmed that this difference only occurs between the AMHA-1 and AMHA-2 cohorts. Meanwhile, the additional post-hoc two-sample K-S test assessment concluded that the distributions for $LA$ across the three cohorts did not differ.

$RA$ represents a function as shown in Figs. 6a-c. The results for $RA$ in Tables 6a & 6b demonstrate that $RA$ distributions across all three cohorts did not differ by failing to reject the null hypothesis for the Kruskal-Wallis $H$-test, as well as the post-hoc Dunn's test and two-sample K-S test assessments.

Loss resilience ($LR$), which is computed the same way as $RA$ using the avoidance curve, and represents the function shown in the third (lower-left) quadrants for Fig. 6a-c. The three rating experiments produced similar functional forms for $LR$ curves; it should be noted that the $LR$ and $RA$ curves with the rating experiment were similar in means and contained overlapping confidence intervals for the mean, as shown in Table 5. Additionally, $LR$ statistically differed across the three cohorts according to Table 6a, however the post-hoc assessments both highlight that for the AMHA-1 and EBS cohorts were not statistically significant ($p > 0.05$).

Positive and negative offsets, $\beta_+$ and $\beta_-$ respectively, are clearly present in both the logarithmic and power law fits to the value function across the three rating experiments. For comparison they were computed from the logarithmic fit and did not significantly differ across cohorts. It should be noted that prospect theory does not allow for offsets to the value function and sets an inflection point connecting the positive and negative arms of the value function so there cannot be offsets for both functions (Kahneman & Tversky, 1979; Tversky & Kahneman, 1992). The current data confirms prior findings using an analysis of RPT features show the existence of clear offsets to the value function from the origin when using $(K, H)$ variables, and a discontinuity along the $K$-axis intercepts between the approach and avoidance data (Kim et al., 2010; Livengood et al., 2017; Viswanathan et al., 2017). Although the $H$-test results in Table 6a fail to reject the null hypothesis for both $\beta_+$ and $\beta_-$, the post-hoc assessments in Table 6b don't demonstrate statistically significant differences in the distributions between AMHA-1 and AMHA-2 for $\beta_+$ and $\beta_-$, and AMHA-1 and EBS for just $\beta_-$.

Apex ($\alpha_\pm$), turning point ($\rho_\pm$), and quadratic area ($QA_\pm$) features are standard metrics of parabolic fits for the limit function (i.e., the fit for the mean-variance curves). The apices ($\alpha_\pm$), statistically differed across the three cohorts, however failed the reject the null hypothesis for some of the post-hoc assessments (see Table 6b). For the positive turning point, $\rho_+$, there was not a statistically significant difference across all cohorts (see Tables 6a & 6b), however the negative turning point $\rho_-$ statistically differed in the Kruskal-Wallis three-way comparison. The post-hoc assessments for $\rho_-$ shows that differences in the distributions were not statistically significant except for the K-S test's results in comparing the AMHA-2 and EBS cohorts; although, Fig. 7 shows overlapping 95% CIs for the median. For the positive quadratic area, $QA_+$, there were not statistically significant differences in approach and avoidance (Tables 6a & 6b), however for $QA_-$ the Kruskal-Wallis comparison (Table 6a) and Table 6b indicates no statistical significance in differences for just the AMHA-1 and EBS cohorts. Qualitatively, the limit curves for approach and avoidance look symmetric to each other relative to the $H$-axis (information/value) for the rating experiment; this is not the case for published keypress data (Kim et al., 2010; Lee et al., 2015; Livengood et al., 2017; Viswanathan et al., 2015, 2017).

Tradeoff curve features showed consistent statistical differences for all four features across the three cohorts for the three-way Kruskal-Wallis comparison (Table 6a), as well as non-overlapping CIs for the mean (Table 5). The mean polar angle, $\theta$, for the rating experiments was $> 45°$, consistent with a slight weighting of the rating assessments toward approach. The consistency between approach and avoidance variables is encoded in the radial distance feature, $r$, (i.e., if the increases in approach balanced decreases in avoidance, or if an individual felt more conflict (increases in both approach and avoidance). The radial distance features for all three experiments were just slightly within the semi-circle described for the $r = \log_2(9)$ or $r = \log_2(8)$, a theoretical frame for ensembles of nine or eight pictures (please see Kim et al., 2010).

# 4   Discussion

Results from this study indicate that:

(1) picture ratings without an operant framework produced RPT curves with the same mathematical form as those produced in an operant context where each action has a consequence by changing the viewing time.

(2) Rating-based RPT curves produced were discrete and recurrent across cohorts and picture ensemble size and scaled from individual to group data.

(3) RPT curves across the three cohorts showed high symmetry between liking and disliking assessment.

(4) Several features of these RPT curves were consistent across the three cohorts, but in some cases, differed relative to other experimental contexts.

As with the operant keypress procedure in other studies, the picture rating task produced data that showed RPT-like relationships (e.g., Fig. 1). The rating task value functions, like those observed with keypressing, followed the pattern observed with prospect theory (Kahneman & Tversky, 1979; Tversky & Kahneman, 1992), and the limit functions followed that described by Markowitz (1952) for risk-reward curves. The $R^2$ across the three studies range from 0.75 to 0.95, in line with previously published results in different cohorts using a keypress paradigm (Kim et al., 2010; Livengood et al., 2017; Viswanathan et al., 2017). Extracted features from the corresponding curves for several of the three rating experiments showed consistent, statistically and visually similar patterns.

The behavioral finance measure of risk aversion ($RA$) did not statistically differ between cohorts, whereas the same metric applied to the avoidance curves (i.e., referred to as $LR$ herein) statistically differed between the AMHA-2 and both the AMHA-1 and EBS cohorts, although the AMHA-1 and EBS cohorts were similar. Altogether, observations point to a greater symmetry in the valuing of positive (approach) and negative (avoidance) aspects of the stimuli when ratings are performed with no behavioral consequence, as opposed to prior studies based on keypressing. Supporting this observation, asymmetries in the mean-variance $(K, \sigma)$ curves observed in prior operant keypress experiments (Livengood et al., 2017; Viswanathan et al., 2017) were not evident in individual or group data from the rating task. These observations with $RA$ and $LR$ support the hypothesis raised by the $LA$ results, suggesting rating-based tasks may reflect less regard for negative consequences.

The $(K, H)$ and $(K, \sigma)$ curves observed for rating tasks were consistent with the model developed by (Robinson & Berridge, 1993) distinguishing "wanting" from "liking", although more work is needed to frame how ratings of "liking" are distinguished from operant keypress procedures that are suggestive of "wanting." In this context it should be noted that keypressing for facial pictures has previously been associated with nucleus accumbens (NAc) activation (Aharon et al., 2001) and $LA$ measures from keypressing (Viswanathan et al., 2015), showing similar responses in the NAc to those of craving (wanting) during cocaine infusions (Breiter et al., 1997).

Although prospect theory (Kahneman & Tversky, 1979) considers the value function to be continuous with an inflection point, the rating task produced offsets, consistent with prior keypress experiments and RPT analyses (Kim et al., 2010; Livengood et al., 2017; Viswanathan et al., 2015, 2017). These offsets are suggestive of other psychological phenomena, such as the ante in poker, where a player must place a bet in the pot to enter the card game (e.g., $\beta_+$), or an insurance premium paid to counter potential bad outcomes (e.g., $\beta_-$). Further work is warranted to frame these findings.

Further research is also needed to deal with caveats to the current work, including that:

(1) demographic matching between cohorts was not perfect, and

(2)  the rating experiment used a unitary Likert-like scale as opposed to two scales wherein approach and avoidance assessments could be assessed independently.

In line with caveat (1), we hypothesize demographic matching between cohorts may have contributed towards the statistical differences observed with the AMHA-2 when compared to AMHA-1 and EBS, such as differences in the distribution of age groups for AMHA-2 compared to the other cohorts. AMHA-2 specifically contains a right-skewed distribution of participants from an older-age group demographic, whereas the distribution of age groups for AMHA-1 and EBS are more uniform.

With replication, the current findings, using an existing panel calibrated to the US Census, contributes to the big data psychology movement where the experimental environment cannot be as well-controlled as in a lab setting, *but* which can be conducted at a large-scale in much shorter time windows and with a major decrease in research team size and experimental time. Large-scale brain imaging and genetic studies are now quite common (e.g., The Connectome Project, ABCD Project, UK Brain Bank) and involve the collection of dense phenotyping data, although these studies are not primarily focused on human psychology and collect data over an extended time window with large human research teams. Studies with Amazon's Mechanical Turk have argued for extension of task-based psychology studies to the web with small research teams (Buhrmester et al., 2015; Casler et al., 2013; Hauser & Schwarz, 2016; Mason & Suri, 2012; Paolacci et al., 2010), although there has been some critique of such practices (Chandler et al., 2019; Cheung et al., 2017; Crump et al., 2013). The current work points to the opportunity for testing computational behavior at larger scales than can be performed in the lab, allowing for greater sampling in the natural variance in measures.

In summary, the results of this study argue, that preference assessments made through ratings can be modelled quantitatively with $R^2$ goodness-of-fits above 0.8, for RPT-based value functions, limit functions, and trade-off functions. These curves appear to differ from those produced from operant keypressing, particularly with the issue of overweighting of perceived negative stimuli relative to positive ones when individuals must trade effort for exposure to the stimulus and may thus have relevance to digital behavior at large.

## Figures

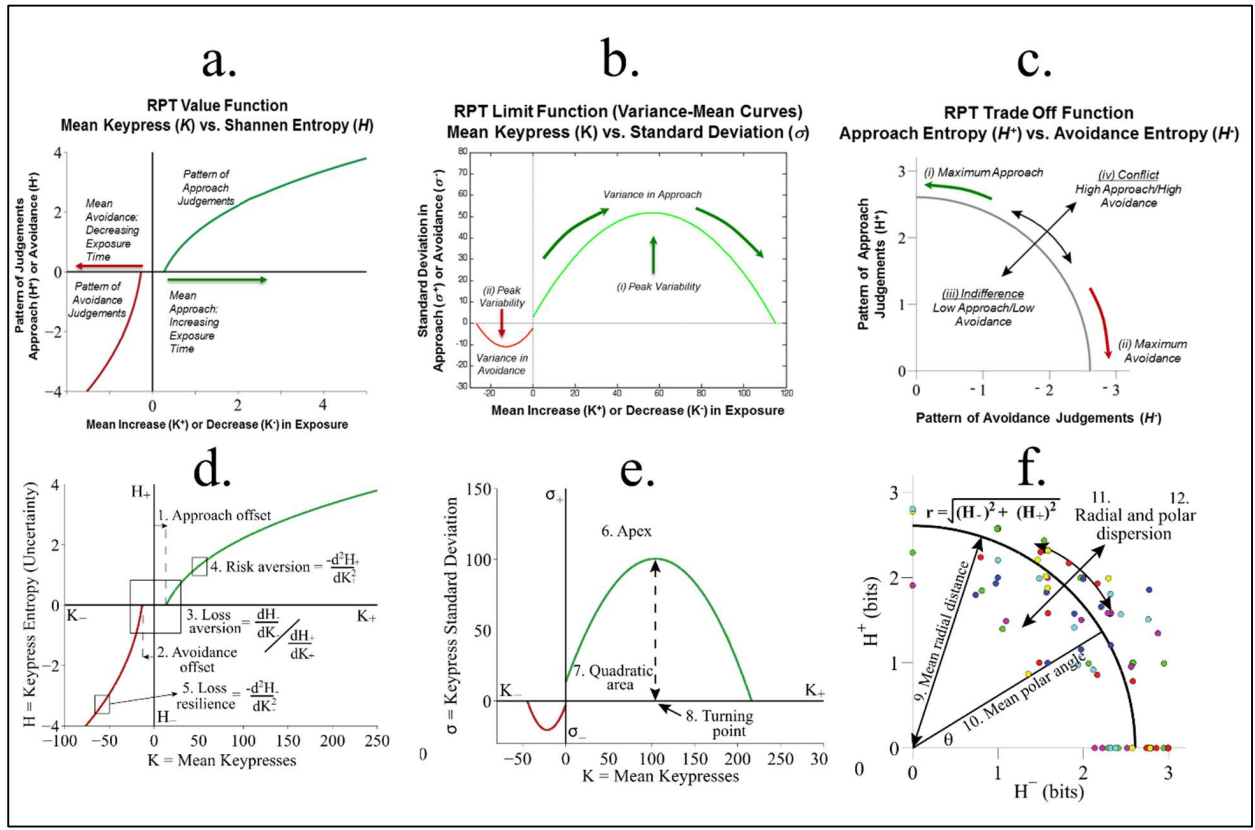

Figure 1. RPT is characterized in part by features that describe relationships between these three sets of behavioral variables: $\{K, H, \sigma\}$ These relationships include: (1) a value function plotting the Shannon entropy ($H_\pm$), against the average ratings ($K_\pm$) for approach or avoidance toward a suite of objects. This function is referred to as a value function given it calibrates "wanting" against the pattern of previous judgements and is consistent with the prospect theory value function. Standard features of these curves, shown in the diagram, include loss aversion ($LA$) and risk aversion ($RA$) from the literature on behavioral economics. The corollary of $RA$ is also shown, herein referred to as loss resilience ($LR$). Two offsets are also noted that are clear in the individual data, relating to an "approach offset" ($\beta_+$) and "avoidance offset" ($\beta_-$). (2) A variance-mean relationship is observed between the average ratings ($K_\pm$) plotted against the corresponding standard deviation of rating responses ($\sigma_\pm$). This relationship is characterized by increasing variance up to a peak followed by decreasing variance back to baseline. This function describes limits to preference or its "saturation" (Fig. 1b). Standard features of this curve include the apices of the quadratic fits, the "turning points" ($\rho_\pm$) or value of $K_\pm$ at which $\sigma_\pm$ is maximal/minimal, and the quadratic areas ($QA_\pm$) of these curves bounded by the $K$-axis. (3) A trade-off function between the approach entropy ($H_+$) and avoidance entropy ($H_-$) was also identified, defining how bundles of approach decisions were balanced with bundles of avoidance decisions as a quantifiable trade-off between approach and avoidance (Fig. 1c). This trade-off function can be characterized by the mean polar angle of the trade-off curves ($\theta$), the standard deviation of this polar angle (its dispersion) ($\sigma_\theta$), the mean radial distance for the trade-off curves ($r$), and its corresponding standard deviation (its dispersion as well) ($\sigma_r$).

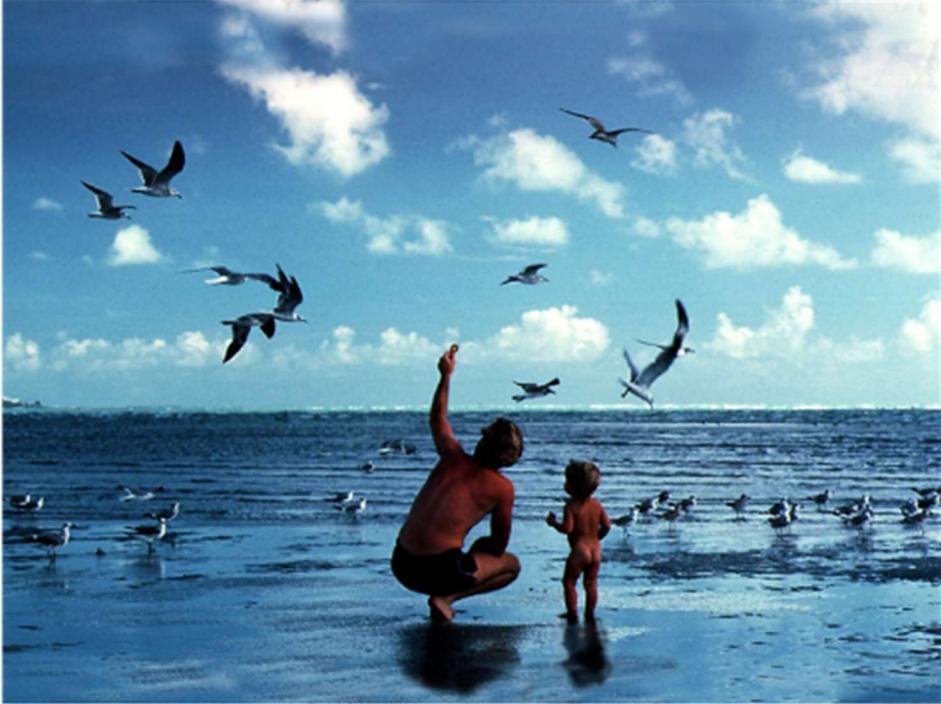

Figure 2. Example of the format of the picture rating task. Unlike the keypress task from prior studies, this task involved no operant consequence to its action. Individuals made ratings with no change in viewing time or other consequence, rating along a 7-point Likert-like scale from −3 to +3.

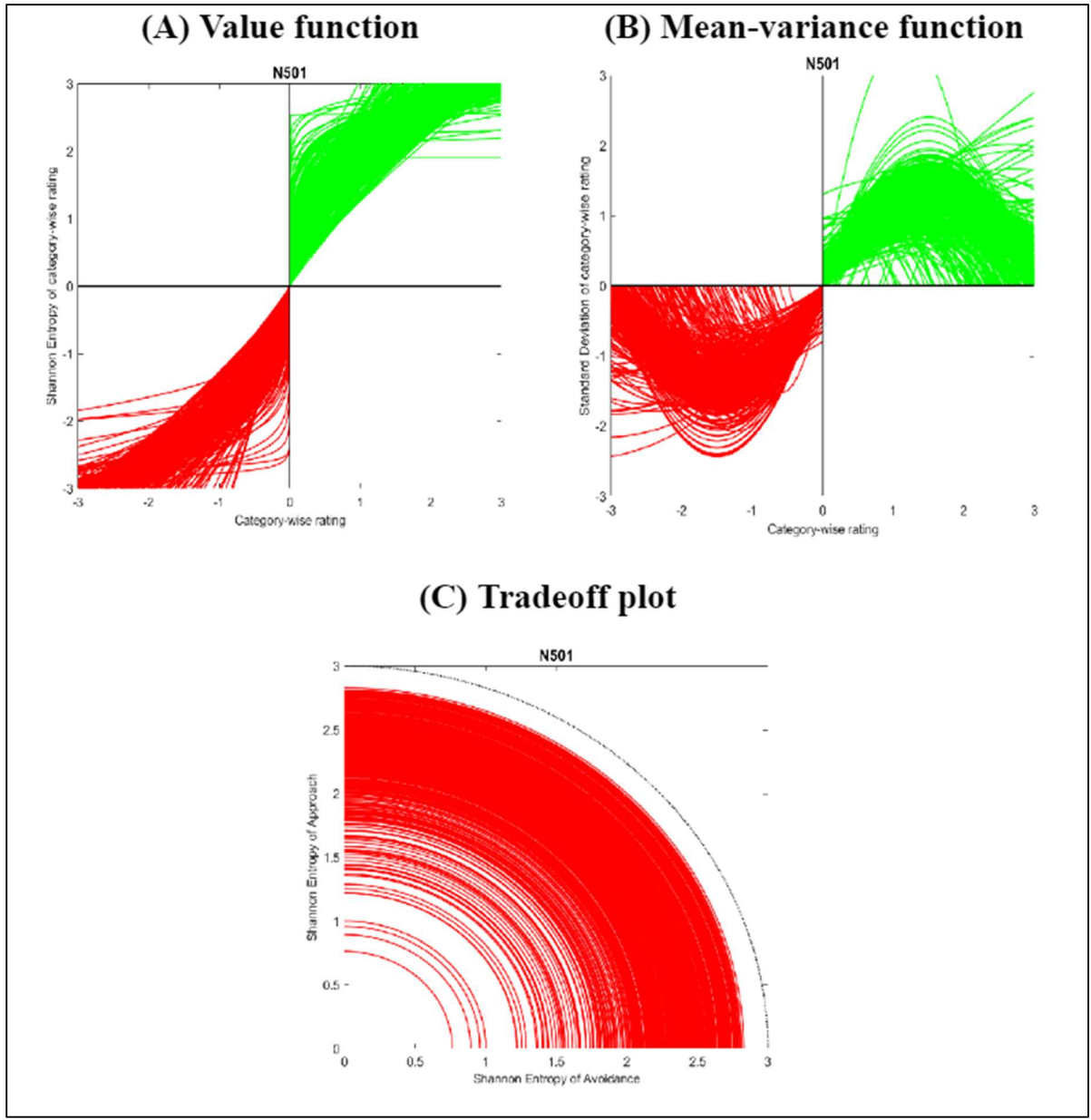

Figure 3. Individual RPT fits of the EBS cohort for the IAPS picture rating task. **(A)** Value functions comparing mean rating intensity ($K$) to rating entropy ($H$) in individual participants. Symbols indicate $K$ and $H$ values computed within six picture categories for either approach ($K_+, H_+$) or avoidance ($K_-, H_-$) rating behavior within a single representative subject. Each color denotes a picture ensemble (i.e., category). Upward-pointing symbols correspond to positive judgements; downward-pointing symbols correspond to negative judgments. Dark green and red traces indicate power-law fits to approach and avoidance data for the representative subject; superimposed light green and red traces indicate individual power-law fits for remaining participants in the cohort. **(B)** Limit functions comparing $K_\pm$ to the standard deviation of approach or avoidance ratings ($\sigma_\pm$) across picture categories in individual participants. Approach and avoidance data for individual participants were fit to quadratic functions (see Methods). **(C)** Trade-off plot comparing entropy for approach ($H_+$) and avoidance ($H_-$) ratings across six picture categories in individual participants. Symbols denote ($H_+, H_-$) data pairs for each picture category in the representative subject. Black line denotes radial fit of ($H_+, H_-$) pairs computed for a representative subject, such that $r = \sqrt{(H_+)^2 + (H_-)^2}$; red lines denote radial fits for remaining participants in cohort.

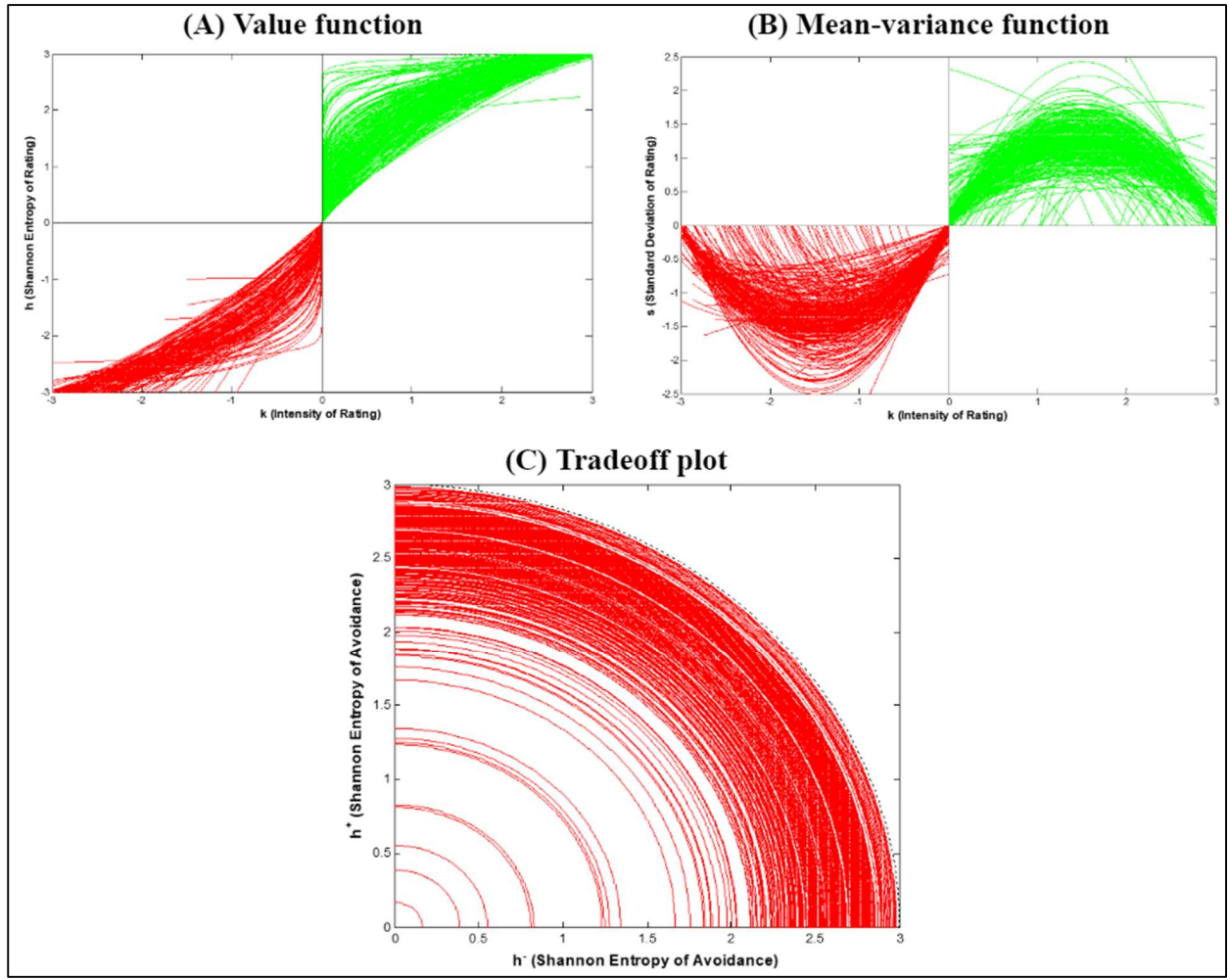

Figure 4. Individual RPT fits of the AMHA-1 cohort for the same IAPS picture rating task used for Fig. 3. **(A)** Value functions comparing mean rating intensity ($K$) to rating entropy ($H$) in individual participants. Please see Fig. 3 for specifics. **(B)** Limit functions comparing $K_{\pm}$ to the standard deviation of approach or avoidance ratings ($\sigma_{\pm}$) across picture categories in individual participants. Please see Fig. 3 for specifics. **(C)** Trade-off plot comparing entropy for approach ($H_+$) and avoidance ($H_-$) ratings across six picture categories in individual participants. Please see Fig. 3 for specifics.

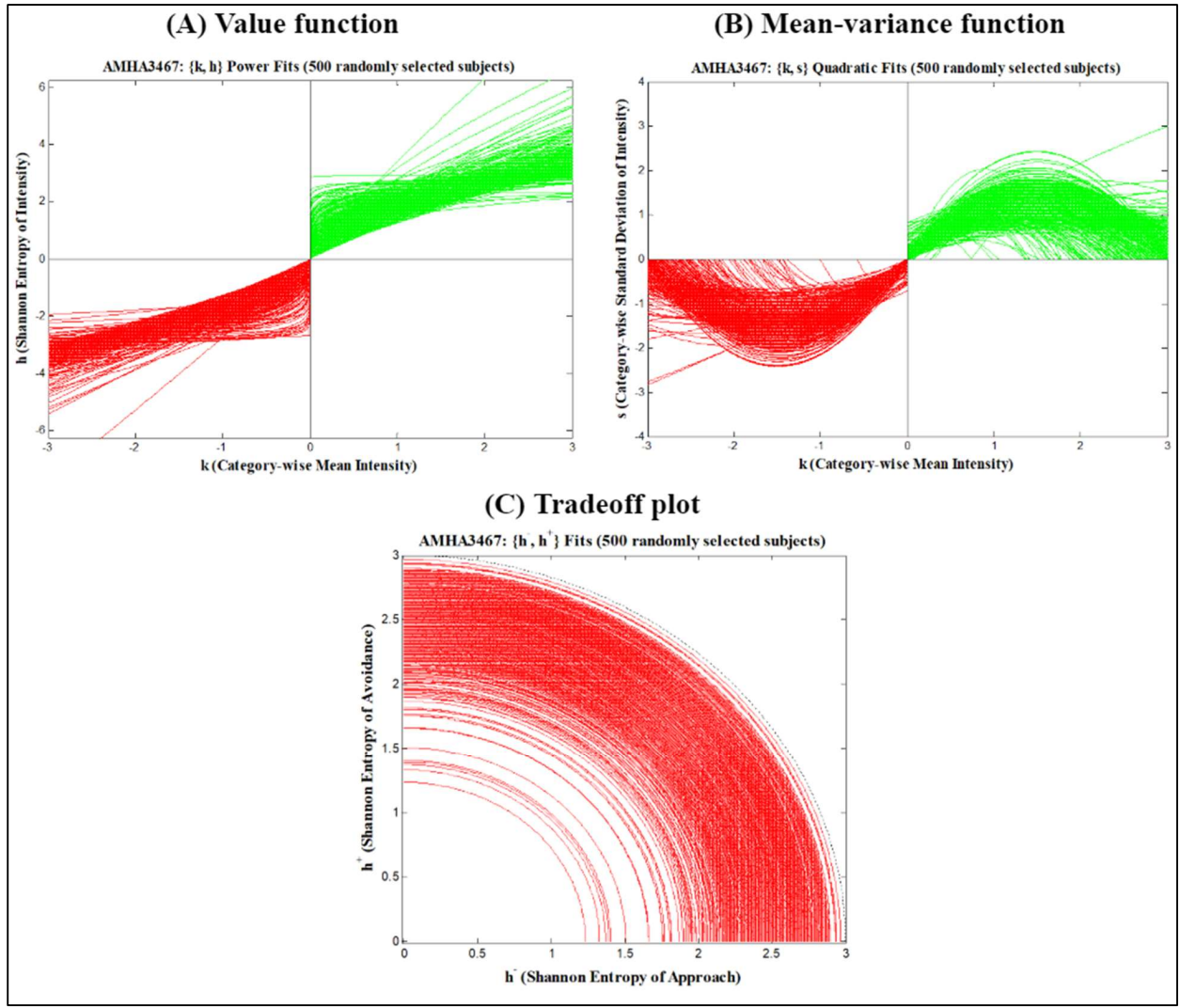

Figure 5. Individual RPT fits of the AMHA-2 cohort for the same IAPS picture rating task used for Fig. 3. **(A)** Value functions comparing mean rating intensity ($K$) to rating entropy ($H$) in individual participants. Please see Fig. 3 for specifics. **(B)** Limit functions comparing $K$ to the standard deviation of approach or avoidance ratings ($\sigma$) across picture categories in individual participants. Please see Fig. 3 for specifics. **(C)** Trade-off plot comparing entropy for approach ($H_+$) and avoidance ($H_-$) ratings across six picture categories in individual participants. Please see Fig. 3 for specifics.

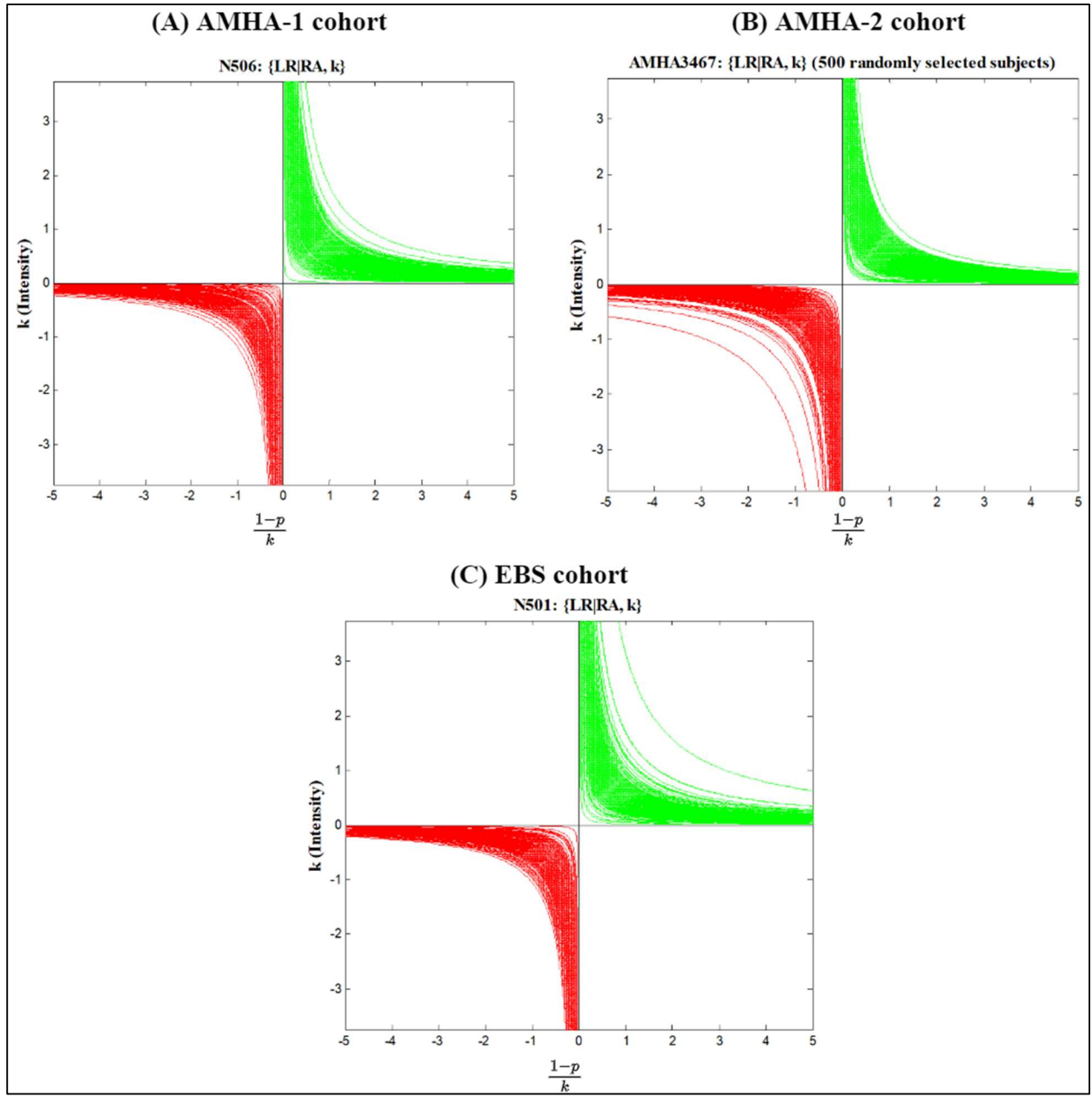

Figure 6. Individual risk aversion ($RA$) and loss resilience ($LR$) functions for the three cohorts. **(A)** Risk aversion ($RA$) functions comparing computed $RA$ to mean picture ratings ($K$) in individual participants are shown for the AMHA-1 cohort. Loss resilience ($LR$) (the same computation as $RA$ done for avoidance ratings) comparing computed $LR$ to mean picture ratings intensity ($K$) in individual participants are shown as well. Note the hyperbolic functional forms in green (approach) and red (avoidance) next to each set of curves. **(B)** Risk aversion functions and loss resilience functions shown for the AMHA-2 cohort. Note the hyperbolic functional forms in green (approach) and red (avoidance) next to each set of curves. **(C)** Risk aversion functions and loss resilience functions show for the EBS cohort. Note the hyperbolic functional forms in green (approach) and red (avoidance) next to each set of curves.

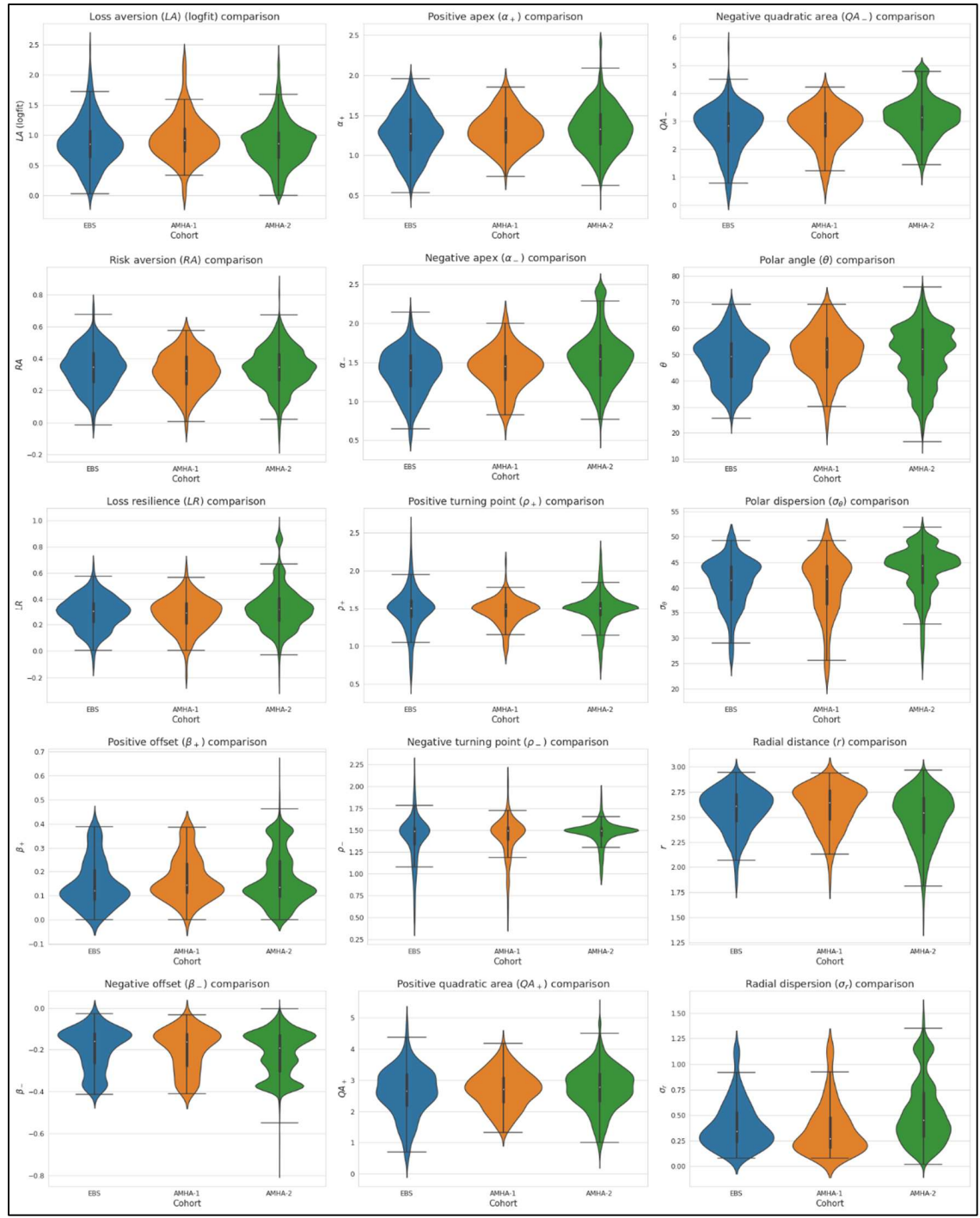

Figure 7. Violin plots (Hintze & Nelson, 1998) for each of the RPT features are tiled to provide a hybrid visual comparison of the distribution, interquartile range (IQR), and 95% CIs, with respect to the corresponding median, for each RPT feature across all cohorts. Kernel density estimates show the shapes of individual distributions for each cohort, while the box plots within each violin plot describes the median and the corresponding IQRs and 95% CIs for the median. Unlike violin plots, box plots are an older standby for visualizing basic distributions which don't allow us to see variations in the data, particularly for multimodal distributions (those with multiple peaks).

# Tables

*Table 1a. Demographics of rating experiment participants in EBS cohort*

| Demographic | Group | Count | Percentage (%) |
|---|---|---|---|
| *Gender identity* | Male | 225 | 44.91 |
| | Female | 276 | 55.09 |
| | Other or prefer not to answer | 0 | 0.00 |
| *Age groups (years)* | 0-17 | 0 | 0.00 |
| | 18-24 | 79 | 15.77 |
| | 25-34 | 107 | 21.36 |
| | 35-44 | 90 | 17.96 |
| | 45-54 | 107 | 21.36 |
| | 55-64 | 102 | 20.36 |
| | 65 ≥ | 16 | 3.19 |
| *Highest education level completed* | Some high school | 19 | 3.79 |
| | High school graduate | 103 | 20.56 |
| | Some college | 201 | 40.12 |
| | Bachelor's degree | 106 | 21.16 |
| | Some graduate school | 9 | 1.80 |
| | Graduate degree | 40 | 7.98 |
| | Post-doctoral training | 23 | 4.59 |
| *Handedness* | Right | 423 | 84.43 |
| | Left | 70 | 13.97 |
| | Both | 8 | 1.60 |
| *Race/Ethnicity* | White/Caucasian | 353 | 70.46 |
| | African American | 80 | 15.97 |
| | Hispanic/Latinx | 20 | 3.99 |
| | Asian or Pacific Islander | 16 | 3.19 |
| | Native American or Alaskan Native | 6 | 1.20 |
| | Mixed racial background | 24 | 4.79 |
| | Other race | 1 | 0.20 |
| | Prefer not to answer | 1 | 0.20 |

*Table 1b. Demographics of rating experiment participants in AMHA-1 cohort*

| Demographic | Group | Count | Percentage (%) |
|---|---|---|---|
| *Gender identity* | Male | 262 | 51.78 |
| | Female | 243 | 48.02 |
| | Other or prefer not to answer | 1 | 0.20 |
| *Age groups (years)* | 0-17 | 0 | 0.00 |
| | 18-24 | 67 | 13.24 |
| | 25-34 | 110 | 21.74 |
| | 35-44 | 128 | 25.30 |
| | 45-54 | 90 | 17.79 |
| | 55-64 | 73 | 14.43 |
| | 65 ≥ | 38 | 7.51 |
| *Highest education level completed* | Some high school | 12 | 2.37 |
| | High school graduate | 114 | 22.53 |
| | Some college | 114 | 22.53 |
| | Bachelor's degree | 114 | 22.53 |
| | Some graduate school | 23 | 4.55 |
| | Graduate degree | 54 | 10.67 |
| | Post-doctoral training | 75 | 14.82 |
| *Handedness* | Right | 393 | 77.67 |
| | Left | 80 | 15.81 |
| | Both | 33 | 6.52 |
| *Race/Ethnicity* | White/Caucasian | 355 | 70.16 |
| | African American | 61 | 12.06 |
| | Hispanic/Latinx | 34 | 6.72 |
| | Asian or Pacific Islander | 15 | 2.96 |
| | Native American or Alaskan Native | 7 | 1.38 |
| | Mixed racial background | 27 | 5.34 |
| | Other race | 4 | 0.79 |
| | Prefer not to answer | 3 | 0.59 |

*Table 1c. Demographics of rating experiment participants in AMHA-2 cohort*

| Demographic | Group | Count | Percentage (%) |
|---|---|---|---|
| Gender identity | Male | 1,592 | 39.61 |
| | Female | 2,408 | 59.92 |
| | Other or prefer not to answer | 19 | 0.47 |
| Age groups (years) | 0-17 | 0 | 0.00 |
| | 18-24 | 245 | 6.10 |
| | 25-34 | 457 | 11.37 |
| | 35-44 | 603 | 15.00 |
| | 45-54 | 702 | 17.47 |
| | 55-64 | 977 | 24.31 |
| | 65 ≥ | 1035 | 25.75 |
| Highest education level completed | Some high school | 112 | 2.79 |
| | High school graduate | 839 | 20.88 |
| | Some college | 1172 | 29.16 |
| | Bachelor's degree | 912 | 22.69 |
| | Some graduate school | 200 | 4.98 |
| | Graduate degree | 666 | 16.57 |
| | Post-doctoral training | 118 | 2.94 |
| Handedness | Right | 3446 | 85.74 |
| | Left | 478 | 11.89 |
| | Both | 95 | 2.36 |
| Race/Ethnicity | White/Caucasian | 3340 | 83.11 |
| | African American | 277 | 6.89 |
| | Hispanic/Latinx | 139 | 3.46 |
| | Asian or Pacific Islander | 145 | 3.61 |
| | Native American or Alaskan Native | 28 | 0.70 |
| | Mixed racial background | 37 | 0.92 |
| | Other race | 21 | 0.52 |
| | Prefer not to answer | 32 | 0.80 |

Legend: Demographic data for subject populations studied by rating task is shown per cohort (Table 1a for EBS; Table 1b for AMHA-1; and Table 1c for AMHA-2. For each demographic, group counts and their relative percentages within the respective cohort are provided.

*Table 2a. Picture ratings summary statistics (EBS cohort)*

| Category | Avoidance | | Approach | |
|---|---|---|---|---|
| | Sum of negative ratings | Mean negative rating per picture | Sum of positive ratings | Mean positive rating per picture |
| Aggressive animals | -6,748 | -2.3354 | 1102 | 1.7225 |
| Nature | -352 | -1.7325 | 7460 | 2.2487 |
| Cute animals | -396 | -1.7071 | 8058 | 2.3523 |
| Disaster | -7203 | -2.4161 | 940 | 1.7608 |
| Nude/exposed bodies | -3092 | -2.1251 | 3529 | 2.1478 |
| Sports | -1700 | -1.8305 | 3657 | 1.7883 |

*Table 2b. Picture ratings summary statistics (AMHA-1 cohort)*

| Category | Avoidance | | Approach | |
|---|---|---|---|---|
| | **Sum of negative ratings** | **Mean negative rating per picture** | **Sum of positive ratings** | **Mean positive rating per picture** |
| *Aggressive animals* | -65,383 | -2.5629 | 6,686 | 1.9861 |
| *Nature* | -3,013 | -1.8483 | 6,0939 | 2.3263 |
| *Cute animals* | -3,723 | -1.8818 | 6,6348 | 2.4463 |
| *Disaster* | -68,694 | -2.5960 | 5,117 | 1.8846 |
| *Nude/exposed bodies* | -23,676 | -2.2657 | 26,998 | 2.1937 |
| *Sports* | -13,971 | -1.9396 | 28,672 | 1.9020 |

*Table 2c. Picture ratings summary statistics (AMHA-2 cohort)*

| Category | Avoidance | | Approach | |
|---|---|---|---|---|
| | **Sum of negative ratings** | **Mean negative rating per picture** | **Sum of positive ratings** | **Mean positive rating per picture** |
| *Aggressive animals* | -4,784 | -2.3719 | 3,024 | 1.9194 |
| *Nature* | -466 | -1.6944 | 7,527 | 2.2516 |
| *Cute animals* | -456 | -1.7757 | 8,111 | 2.3284 |
| *Disaster* | -5255 | -2.4392 | 2,904 | 1.9926 |
| *Nude/exposed bodies* | -1697 | -2.1025 | 5,227 | 2.0948 |
| *Sports* | -1298 | -1.8180 | 5,248 | 1.9844 |

Legend: Summary and descriptive statistics for the "total sum responses across the cohort" and "mean response", stratified by positive or negative valence of ratings (approach, $+$, or avoidance $-$). The total sum of approach ($+$) and avoidance ($-$) responses is shown per cohort (Table 2a for EBS; Table 2b for AMHA-1; and Table 2c for AMHA-2) for each category of picture shown from the IAPS stimulus set, along with the average response for each picture, per category throughout Tables 2a-2c.

*Table 3a. Group fitting parameters for rating experiment of EBS cohort*

| Curve set | Curve | $R^2$ | Parameters |
|---|---|---|---|
| $(K, H)$ **power law** | $(K_+, H_+)$ | 0.732 | $\begin{cases} a = 0.440 \\ b = 0.296 \end{cases}$ |
| | $(K_-, H_-)$ | 0.766 | $\begin{cases} a = 0.473 \\ b = 0.274 \end{cases}$ |
| $(\log K, H)$ **logarithmic** | $(\log K_+, H_+)$ | 0.857 | $\begin{cases} a = 2.210 \\ b = 2.047 \end{cases}$ |
| | $(\log K_-, H_-)$ | 0.871 | $\begin{cases} a = 2.330 \\ b = 1.932 \end{cases}$ |
| $(K, \sigma)$ **quadratic** | $(K_+, \sigma_+)$ | 0.750 | $\begin{cases} a = -4.755 \times 10^{-1} \\ b = 1.468 \\ c = 0.067 \end{cases}$ |
| | $(K_-, \sigma_-)$ | 0.817 | $\begin{cases} a = -5.530 \times 10^{-1} \\ b = 1.667 \\ c = 0.065 \end{cases}$ |

*Table 3b. Group fitting parameters for rating experiment of AMHA-1 cohort*

| Curve set | Curve | $R^2$ | Parameters |
|---|---|---|---|
| $(K, H)$ **power law** | $(K_+, H_+)$ | 0.784 | $\begin{cases} a = 0.475 \\ b = 0.287 \end{cases}$ |
| | $(K_-, H_-)$ | 0.809 | $\begin{cases} a = 0.502 \\ b = 0.261 \end{cases}$ |
| $(\log K, H)$ **logarithmic** | $(\log K_+, H_+)$ | 0.890 | $\begin{cases} a = 2.316 \\ b = 2.013 \end{cases}$ |
| | $(\log K_-, H_-)$ | 0.892 | $\begin{cases} a = 2.324 \\ b = 1.905 \end{cases}$ |
| $(K, \sigma)$ **quadratic** | $(K_+, \sigma_+)$ | 0.746 | $\begin{cases} a = -4.702 \times 10^{-1} \\ b = 1.439 \\ c = 0.091 \end{cases}$ |
| | $(K_-, \sigma_-)$ | 0.862 | $\begin{cases} a = -5.804 \times 10^{-1} \\ b = 1.754 \\ c = 0.055 \end{cases}$ |

*Table 3c. Group fitting parameters for rating experiment of AMHA-2 cohort*

| Curve set | Curve | $R^2$ | Parameters |
|---|---|---|---|
| $(K, H)$ **power law** | $(K_+, H_+)$ | 0.767 | $\begin{cases} a = 0.461 \\ b = 0.282 \end{cases}$ |
| | $(K_-, H_-)$ | 0.827 | $\begin{cases} a = 0.510 \\ b = 0.250 \end{cases}$ |
| $(\log K, H)$ **logarithmic** | $(\log K_+, H_+)$ | 0.875 | $\begin{cases} a = 2.304 \\ b = 1.967 \end{cases}$ |
| | $(\log K_-, H_-)$ | 0.902 | $\begin{cases} a = 2.394 \\ b = 1.851 \end{cases}$ |
| $(K, \sigma)$ **quadratic** | $(K_+, \sigma_+)$ | 0.812 | $\begin{cases} a = -5.182 \times 10^{-1} \\ b = 1.588 \\ c = 0.068 \end{cases}$ |
| | $(K_-, \sigma_-)$ | 0.864 | $\begin{cases} a = -6.087 \times 10^{-1} \\ b = 1.827 \\ c = 0.069 \end{cases}$ |

Legend: Group power law, $(K, H)$, and logarithmic, $(\log K, H)$ fits along with quadratic, $(K, \sigma)$, fits are listed for the rating data across the three cohorts, stratified by valence of ratings (approach, $+$; or avoidance $-$). For each fit, the coefficient of determination (goodness of fit), $R^2$, and the corresponding fitting parameters of the group approach $(+)$ and avoidance $(-)$ curves is reported per cohort (Table 3a for EBS; Table 3b for AMHA-1; and Table 3c for AMHA-2). The coefficient of determination, $R^2$, shows the percent of the variation in each fit that is explained by the model.

*Table 4. Goodness of fit summary statistics for individual value and limit functions*

| Curve Set | Curve | Summary Statistic | Mean ± SD (per cohort) | | |
|---|---|---|---|---|---|
| | | | *EBS* | *AMHA-1* | *AMHA-2* |
| $(\log K, H)$ | $(\log K_+, H_+)$ | $R^2$ | $0.89 \pm 0.14$ | $0.89 \pm 0.16$ | $0.92 \pm 0.13$ |
| | | adjusted $R^2$ | $0.85 \pm 0.21$ | $0.85 \pm 0.23$ | $0.87 \pm 0.21$ |
| | | $F$-value | $1.23 \times 10^{30}$ $\pm 1.01 \times 10^{31}$ | $8.00 \times 10^{29}$ $\pm 1.24 \times 10^{31}$ | $1.90 \times 10^{30}$ $\pm 2.06 \times 10^{31}$ |
| | $(\log K_-, H_-)$ | $R^2$ | $0.94 \pm 0.11$ | $0.92 \pm 0.12$ | $0.96 \pm 0.07$ |
| | | adjusted $R^2$ | $0.90 \pm 0.18$ | $0.88 \pm 0.21$ | $0.80 \pm 0.29$ |
| | | $F$-value | $2.44 \times 10^{30}$ $\pm 2.21 \times 10^{31}$ | $1.85 \times 10^{30}$ $\pm 1.43 \times 10^{31}$ | $3.61 \times 10^{30}$ $\pm 2.34 \times 10^{31}$ |
| $(\log K, \log H)$ | $(\log K_+, \log H_+)$ | $R^2$ | $0.84 \pm 0.20$ | $0.86 \pm 0.17$ | $0.89 \pm 0.17$ |
| | | adjusted $R^2$ | $0.76 \pm 0.32$ | $0.80 \pm 0.26$ | $0.80 \pm 0.29$ |
| | | $F$-value | $2.21 \times 10^{29}$ $\pm 2.20 \times 10^{30}$ | $1.10 \times 10^{30}$ $\pm 1.39 \times 10^{31}$ | $3.43 \times 10^{29}$ $\pm 6.57 \times 10^{30}$ |
| | $(\log K_-, \log H_-)$ | $R^2$ | $0.92 \pm 0.12$ | $0.91 \pm 0.14$ | $0.95 \pm 0.10$ |
| | | adjusted $R^2$ | $0.87 \pm 0.21$ | $0.85 \pm 0.24$ | $0.90 \pm 0.20$ |
| | | $F$-value | $1.40 \times 10^{30}$ $\pm 1.26 \times 10^{31}$ | $7.15 \times 10^{29}$ $\pm 4.59 \times 10^{30}$ | $1.14 \times 10^{30}$ $\pm 8.26 \times 10^{30}$ |
| $(K, \sigma)$ | $(K_+, \sigma_+)$ | $R^2$ | $0.87 \pm 0.18$ | $0.85 \pm 0.19$ | $0.90 \pm 0.13$ |
| | | adjusted $R^2$ | $0.78 \pm 0.30$ | $0.75 \pm 0.31$ | $0.84 \pm 0.21$ |
| | | $F$-value | $1.24 \times 10^{30}$ $\pm 1.52 \times 10^{31}$ | $2.33 \times 10^{32}$ $\pm 3.95 \times 10^{33}$ | $1.12 \times 10^{30}$ $\pm 1.50 \times 10^{31}$ |
| | $(K_-, \sigma_-)$ | $R^2$ | $0.93 \pm 0.11$ | $0.93 \pm 0.10$ | $0.94 \pm 0.10$ |
| | | adjusted $R^2$ | $0.88 \pm 0.18$ | $0.89 \pm 0.17$ | $0.90 \pm 0.17$ |
| | | $F$-value | $1.34 \times 10^{31}$ $\pm 1.52 \times 10^{32}$ | $6.93 \times 10^{31}$ $\pm 9.46 \times 10^{32}$ | $2.26 \times 10^{30}$ $\pm 5.00 \times 10^{31}$ |

Legend: Individual logarithmic, $(\log K, H)$, and linear, $(\log K, \log H)$, fits, along with quadratic, $(K, \sigma)$, fits are listed for the rating data across the three cohorts. Linear, logarithmic, and quadratic correlations were performed in each subject across the data relating to approach ratings for the six categories of IAPS stimuli, and across the data relating to avoidance responses; participants needed data from at least two of the experimental conditions (aggressive animals, nature, cute animals, disaster, nudity, and sports) to be fitted. The mean and standard deviation (SD) are listed for the coefficient of determination, $R^2$, the corresponding adjusted value, $R^2_{adj}$, and corresponding $F$-test values.

*Table 5. RPT curve metrics for the rating experiments across the three cohorts*

| RPT metric | Statistic | Cohort | | |
|---|---|---|---|---|
| | | **EBS** | **AMHA-1** | **AMHA-2** |
| Loss aversion (logfit) (LA) | Mean ± SD | 0.88 ± 1.148 | 1.51 ± 2.090 | 0.88 ± 0.365 |
| | SEM | 0.0738 | 0.1647 | 0.0063 |
| | 95% CI | [0.74, 1.03] | [1.19, 1.84] | [0.86, 0.89] |
| Risk aversion (RA) | Mean ± SD | 0.35 ± 0.122 | 0.32 ± 0.120 | 0.34 ± 0.125 |
| | SEM | 0.0072 | 0.0090 | 0.0022 |
| | 95% CI | [0.33, 0.36] | [0.31, 0.34] | [0.34, 0.35] |
| Loss resilience (LR) | Mean ± SD | 0.30 ± 0.122 | 0.30 ± 0.130 | 0.32 ± 0.134 |
| | SEM | 0.0073 | 0.0098 | 0.0024 |
| | 95% CI | [0.29, 0.32] | [0.28, 0.31] | [0.32, 0.33] |
| Positive offset ($\beta_+$)* | Mean ± SD | 0.15 ± 0.101 | 0.16 ± 0.097 | 0.17 ± 0.106 |
| | SEM | 0.0046 | 0.0052 | 0.0018 |
| | 95% CI | [0.14, 0.16] | [0.15, 0.17] | [0.17, 0.18] |
| Negative offset ($\beta_-$)* | Mean ± SD | -0.19 ± 0.101 | -0.19 ± 0.101 | -0.21 ± 0.108 |
| | SEM | 0.0048 | 0.0061 | 0.0018 |
| | 95% CI | [-0.20, -0.18] | [-0.20, -0.18] | [-0.21, -0.21] |
| Positive apex ($\alpha_+$)* | Mean ± SD | 1.27 ± 0.327 | 1.27 ± 0.274 | 1.33 ± 0.290 |
| | SEM | 0.0154 | 0.0152 | 0.0049 |
| | 95% CI | [1.24, 1.30] | [1.24, 1.30] | [1.32, 1.34] |
| Negative apex ($\alpha_-$)* | Mean ± SD | 1.39 ± 0.378 | 1.40 ± 0.382 | 1.52 ± 0.349 |
| | SEM | 0.0174 | 0.0216 | 0.0059 |
| | 95% CI | [1.35, 1.42] | [1.35, 1.44] | [1.51, 1.53] |
| Positive turning point ($\rho_+$)* | Mean ± SD | 1.48 ± 0.269 | 1.48 ± 0.203 | 1.48 ± 0.195 |
| | SEM | 0.0131 | 0.0116 | 0.0034 |
| | 95% CI | [1.45, 1.50] | [1.45, 1.50] | [1.47, 1.48] |
| Negative turning point ($\rho_-$)* | Mean ± SD | 1.40 ± 0.275 | 1.38 ± 0.301 | 1.48 ± 0.103 |
| | SEM | 0.0131 | 0.0176 | 0.0019 |
| | 95% CI | [1.37, 1.42] | [1.34, 1.41] | [1.48, 1.49] |
| Positive quadratic area ($QA_+$)* | Mean ± SD | 2.63 ± 0.867 | 2.59 ± 0.783 | 2.71 ± 0.759 |
| | SEM | 0.0412 | 0.0437 | 0.0130 |
| | 95% CI | [2.55, 2.71] | [2.51, 2.68] | [2.68, 2.73] |
| Negative quadratic area ($QA_-$)* | Mean ± SD | 2.68 ± 1.031 | 2.66 ± 1.028 | 3.05 ± 0.815 |
| | SEM | 0.0477 | 0.0583 | 0.0139 |
| | 95% CI | [2.58, 2.77] | [2.55, 2.78] | [3.02, 3.08] |
| Polar angle ($\theta$) | Mean ± SD | 52.53 ± 14.654 | 58.89 ± 16.927 | 50.04 ± 11.499 |
| | SEM | 0.6560 | 0.8872 | 0.1950 |
| | 95% CI | [51.24, 53.82] | [57.14, 60.63] | [49.66, 50.42] |
| Polar dispersion ($\sigma_\theta$) | Mean ± SD | 40.73 ± 7.024 | 34.40 ± 15.146 | 40.73 ± 7.024 |
| | SEM | 0.3203 | 0.7917 | 0.3203 |
| | 95% CI | [40.10, 41.36] | [32.84, 35.95] | [40.10, 41.36] |
| Radial distance (r) | Mean ± SD | 2.52 ± 0.256 | 2.60 ± 0.239 | 2.52 ± 0.256 |
| | SEM | 0.0115 | 0.0127 | 0.0115 |
| | 95% CI | [2.50, 2.55] | [2.57, 2.62] | [2.50, 2.55] |
| Radial dispersion ($\sigma_r$) | Mean ± SD | 0.47 ± 0.298 | 0.38 ± 0.276 | 0.47 ± 0.298 |
| | SEM | 0.0133 | 0.0145 | 0.0133 |
| | 95% CI | [0.44, 0.49] | [0.35, 0.40] | [0.44, 0.49] |

Legend: Descriptive statistics for features of the $(K, H)$, $(K, \sigma)$, and $(H_-, H_+)$ curves of the IAPS picture rating data across the three distinct cohorts. Fifteen features were identified using common engineering methods, including five features from the value function, six from the limit function, and four from the trade-off function (see Methods). For each of the three datasets, the mean and standard deviation (SD) are listed for the fifteen features, along with standard error of the mean (SEM) and the 95% confidence intervals (CI) for the corresponding means.

*Table 6a. Nonparametric comparison of RPT metrics across the three distinct cohorts for IAPS picture rating experiment*

| RPT metric | Kruskal-Wallis $H$-test (degrees of freedom = 2) | |
|---|---|---|
| | *H-value* | *p-value* |
| *Loss aversion (logfit) (LA)* | 6.98785 | $3.0381 \times 10^{-2}$ |
| *Risk aversion (RA)* | 4.07003 | $1.3068 \times 10^{-1}$ |
| *Loss resilience (LR)* | 13.3089 | $1.2883 \times 10^{-3}$ |
| *Positive offset ($\beta_+$)* | 10.2435 | $5.9655 \times 10^{-3}$ |
| *Negative offset ($\beta_-$)* | 20.1216 | $4.2722 \times 10^{-5}$ |
| *Positive apex ($\alpha_+$)* | 14.3479 | $7.6627 \times 10^{-4}$ |
| *Negative apex ($\alpha_-$)* | 74.5914 | $6.3486 \times 10^{-17}$ |
| *Positive turning point ($\rho_+$)* | 2.28294 | $3.1935 \times 10^{-1}$ |
| *Negative turning point ($\rho_-$)* | 6.16587 | $4.5824 \times 10^{-2}$ |
| *Positive quadratic area ($QA_+$)* | 5.38677 | $6.7652 \times 10^{-2}$ |
| *Negative quadratic area ($QA_-$)* | 65.4338 | $6.1835 \times 10^{-15}$ |
| *Polar angle ($\theta$)* | 9.9913 | $6.7672 \times 10^{-3}$ |
| *Polar dispersion ($\sigma_\theta$)* | 125.001 | $7.1836 \times 10^{-28}$ |
| *Radial distance ($r$)* | 40.4942 | $1.6099 \times 10^{-9}$ |
| *Radial dispersion ($\sigma_r$)* | 89.6117 | $3.4759 \times 10^{-20}$ |

*Table 6b. Corresponding post-hoc pairwise Dunn's test and two-sample Kolmogorov-Smirnov (K-S) test statistics*

| RPT metric | Dunn's test $p$-values (Holm-Bonferroni corrected) | | | Two-sample K-S test $p$-values | | |
|---|---|---|---|---|---|---|
| | *AMHA-1/AMHA-2* | *AMHA-1/EBS* | *EBS/AMHA-2* | *AMHA-1/AMHA-2* | *AMHA-1/EBS* | *EBS/AMHA-2* |
| *Loss aversion (logfit) (LA)* | 0.025 | 0.097 | 0.800 | 0.051 | 0.150 | 0.630 |
| *Risk aversion (RA)* | 0.130 | 0.290 | 0.790 | 0.052 | 0.140 | 0.790 |
| *Loss resilience (LR)* | 0.019 | 0.590 | 0.019 | 0.017 | 0.840 | 0.027 |
| *Positive offset ($\beta_+$)* | 0.400 | 0.017 | $7.50 \times 10^{-3}$ | 0.190 | $8.90 \times 10^{-4}$ | 0.001 |
| *Negative offset ($\beta_-$)* | 0.078 | 0.330 | $1.20 \times 10^{-4}$ | 0.077 | 0.500 | $1.10 \times 10^{-3}$ |
| *Positive apex ($\alpha_+$)* | 0.580 | 0.094 | $4.80 \times 10^{-4}$ | 0.270 | 0.015 | $1.70 \times 10^{-3}$ |
| *Negative apex ($\alpha_-$)* | $0.400 \times 10^{-6}$ | 0.320 | $2.00 \times 10^{-13}$ | $4.300 \times 10^{-6}$ | 0.360 | $1.10 \times 10^{-9}$ |
| *Positive turning point ($\rho_+$)* | 0.420 | 0.420 | 0.820 | 0.099 | 0.057 | 0.490 |
| *Negative turning point ($\rho_-$)* | 0.670 | 0.670 | 0.057 | 0.061 | 0.620 | $2.90 \times 10^{-4}$ |
| *Positive quadratic area ($QA_+$)* | 0.790 | 0.790 | 0.082 | 0.380 | 0.410 | 0.073 |
| *Negative quadratic area ($QA_-$)* | $1.10 \times 10^{-5}$ | 0.410 | $9.60 \times 10^{-12}$ | $3.60 \times 10^{-4}$ | 0.640 | $1.60 \times 10^{-7}$ |
| *Polar angle ($\theta$)* | 0.340 | 0.015 | $9.60 \times 10^{-3}$ | 0.018 | 0.026 | $5.30 \times 10^{-6}$ |
| *Polar dispersion ($\sigma_\theta$)* | $6.00 \times 10^{-13}$ | 0.890 | $1.40 \times 10^{-18}$ | $2.10 \times 10^{-11}$ | 0.560 | $3.80 \times 10^{-15}$ |
| *Radial distance ($r$)* | $1.500 \times 10^{-6}$ | 0.180 | $5.10 \times 10^{-5}$ | $1.30 \times 10^{-5}$ | 0.055 | $2.90 \times 10^{-4}$ |
| *Radial dispersion ($\sigma_r$)* | $9.600 \times 10^{-13}$ | 0.090 | $1.70 \times 10^{-10}$ | $1.10 \times 10^{-11}$ | $8.10 \times 10^{-4}$ | $1.60 \times 10^{-9}$ |

Legend: Non-parametric statistics comparing the fifteen RPT features listed in Table 5 across the three cohorts for the rating task experiment. Table 6a provides the Kruskal-Wallis $H$-test scores and their corresponding $p$-values for a three-way comparison in determining statistically significant differences in the distributions across the three rating task cohorts. Given at least 5 observations for each group, of three total, we can approximate the critical values regarding the null hypothesis using the same critical values as a $X^2$-test, given two degrees of freedom for this comparison (number of groups minus 1). As a post-hoc assessment, Table 6b contains the $p$-values from a pairwise comparison between cohorts using Dunn's test ($p$-values are corrected using the Holm-Bonferroni method), as well as $p$-values from multiple pairwise comparisons of the distributions for each RPT metric across the cohorts using two-sample Kolmogorov-Smirnov (K-S) nonparametric tests.

# References

Aharon, I., Etcoff, N., Ariely, D., Chabris, C. F., O'Connor, E., & Breiter, H. C. (2001). Beautiful Faces Have Variable Reward Value: fMRI and Behavioral Evidence. *Neuron*, *32*(3), 537–551. https://doi.org/10.1016/S0896-6273(01)00491-3

Banks, H. T., & Tran, H. T. (2009). *Mathematical and experimental modeling of physical and biological processes*. CRC Press.

Berridge, K. C., & Robinson, T. E. (2016). The Mind of an Addicted Brain: Neural Sensitization of Wanting Versus Liking: *Https://Doi.Org/10.1111/1467-8721.Ep10772316*, *4*(3), 71–75. https://doi.org/10.1111/1467-8721.EP10772316

Breiter, H. C., Gollub, R. L., Weisskoff, R. M., Kennedy, D. N., Makris, N., Berke, J. D., Goodman, J. M., Kantor, H. L., Gastfriend, D. R., Riorden, J. P., Mathew, R. T., Rosen, B. R., & Hyman, S. E. (1997). Acute effects of cocaine on human brain activity and emotion. *Neuron*, *19*(3), 591–611. https://doi.org/10.1016/S0896-6273(00)80374-8

Breiter, H. C., & Kim, B. W. (2008). Recurrent and robust patterns underlying human relative preference and associations with brain circuitry plus genetics. *Design Principles in Biology (University of Minnesota, Institute of Mathematics and Its Applications)*, *4*, 21–25.

Buhrmester, M., Kwang, T., & Gosling, S. D. (2015). Amazon's Mechanical Turk: A new source of inexpensive, yet high-quality data? *Methodological Issues and Strategies in Clinical Research (4th Ed.)*, 133–139. https://doi.org/10.1037/14805-009

Casler, K., Bickel, L., & Hackett, E. (2013). Separate but equal? A comparison of participants and data gathered via Amazon's MTurk, social media, and face-to-face behavioral testing. *Computers in Human Behavior*, *29*(6), 2156–2160. https://doi.org/10.1016/j.chb.2013.05.009

Chandler, J., Rosenzweig, C., Moss, A. J., Robinson, J., & Litman, L. (2019). Online panels in social science research: Expanding sampling methods beyond Mechanical Turk. *Behavior Research Methods*, *51*(5), 2022–2038. https://doi.org/10.3758/S13428-019-01273-7

Cheung, J. H., Burns, D. K., Sinclair, R. R., & Sliter, M. (2017). Amazon Mechanical Turk in Organizational Psychology: An Evaluation and Practical Recommendations. *Journal of Business and Psychology*, *32*(4), 347–361. https://doi.org/10.1007/S10869-016-9458-5

Crump, M. J. C., McDonnell, J. v., & Gureckis, T. M. (2013). Evaluating Amazon's Mechanical Turk as a Tool for Experimental Behavioral Research. *PLOS ONE*, *8*(3), e57410. https://doi.org/10.1371/JOURNAL.PONE.0057410

Dennis, M. L., Chan, Y. F., & Funk, R. R. (2006). Development and Validation of the GAIN Short Screener (GSS) for Internalizing, Externalizing and Substance Use Disorders and Crime/Violence Problems Among Adolescents and Adults. *The American Journal on Addictions*, *15*(SUPPL. 1), s80–s91. https://doi.org/10.1080/10550490601006055

Dinno, A. (2015). Nonparametric Pairwise Multiple Comparisons in Independent Groups using Dunn's Test: *Https://Doi.Org/10.1177/1536867X1501500117*, *15*(1), 292–300. https://doi.org/10.1177/1536867X1501500117

Elman, I., Ariely, D., Mazar, N., Aharon, I., Lasko, N. B., Macklin, M. L., Orr, S. P., Lukas, S. E., & Pitman, R. K. (2005). Probing reward function in post-traumatic stress disorder with beautiful facial images. *Psychiatry Research*, *135*(3), 179–183. https://doi.org/10.1016/J.PSYCHRES.2005.04.002

Feynman, R. P. (1965). *The Character of Physical Law*. British Broadcasting Corporation.

Gasic, G. P., Smoller, J. W., Perlis, R. H., Sun, M., Lee, S., Kim, B. W., Lee, M. J., Holt, D. J., Blood, A. J., Makris, N., Kennedy, D. K., Hoge, R. D., Calhoun, J., Fava, M., Gusella, J. F., & Breiter, H. C. (2009). BDNF, relative preference, and reward circuitry responses to emotional communication. *American Journal of Medical Genetics, Part B: Neuropsychiatric Genetics*, *150*(6), 762–781. https://doi.org/10.1002/AJMG.B.30944

Hauser, D. J., & Schwarz, N. (2016). Attentive Turkers: MTurk participants perform better on online attention checks than do subject pool participants. *Behavior Research Methods*, *48*(1), 400–407. https://doi.org/10.3758/S13428-015-0578-Z/TABLES/1

Hintze, J. L., & Nelson, R. D. (1998). Violin Plots: A Box Plot-Density Trace Synergism. *The American Statistician*, *52*(2), 181. https://doi.org/10.2307/2685478

Kahneman, D., & Tversky, A. (1979). Prospect Theory: An Analysis of Decision under Risk. *Econometrica*, *47*(2), 263. https://doi.org/10.2307/1914185

Kendall, K., & George, M. (2008a). Kolmogorov–Smirnov Test. *The Concise Encyclopedia of Statistics*, 283–287. https://doi.org/10.1007/978-0-387-32833-1_214

Kendall, K., & George, M. (2008b). Kruskal-Wallis Test. *The Concise Encyclopedia of Statistics*, 288–290. https://doi.org/10.1007/978-0-387-32833-1_216

Kim, B. W., Kennedy, D. N., Lehár, J., Lee, M. J., Blood, A. J., Lee, S., Perlis, R. H., Smoller, J. W., Morris, R., Fava, M., & Breiter, H. C. (2010). Recurrent, Robust and Scalable Patterns Underlie Human Approach and Avoidance. *PLOS ONE*, *5*(5), e10613. https://doi.org/10.1371/JOURNAL.PONE.0010613

Kroenke, K., Spitzer, R. L., & Williams, J. B. W. (2001). The PHQ-9: validity of a brief depression severity measure. *Journal of General Internal Medicine*, *16*(9), 606–613. https://doi.org/10.1046/J.1525-1497.2001.016009606.X

Lang, P. J., Bradley, M. M., & Cuthbert, B. N. (2008). *International affective picture system (IAPS): Affective ratings of pictures and instruction manual. Technical Report A-8*.

Lee, S., Lee, M. J., Kim, B. W., Gilman, J. M., Kuster, J. K., Blood, A. J., Kuhnen, C. M., & Breiter, H. C. (2015). The Commonality of Loss Aversion across Procedures and Stimuli. *PLOS ONE*, *10*(9), e0135216. https://doi.org/10.1371/JOURNAL.PONE.0135216

Levy, B., Ariely, D., Mazar, N., Chi, W., Lukas, S., & Elman, I. (2008). Gender differences in the motivational processing of facial beauty. *Learning and Motivation*, *39*(2), 136–145. https://doi.org/10.1016/J.LMOT.2007.09.002

Lewin, K. (1935). A dynamic theory of personality. *Development, Factor Analysis, and Validation*.

Livengood, S. L., Sheppard, J. P., Kim, B. W., Malthouse, E. C., Bourne, J. E., Barlow, A. E., Lee, M. J., Marin, V., O'Connor, K. P., Csernansky, J. G., Block, M. P., Blood, A. J., & Breiter, H. C. (2017).

Keypress-based musical preference is both individual and lawful. *Frontiers in Neuroscience*, *11*, 136. https://doi.org/10.3389/FNINS.2017.00136/BIBTEX

Makris, N., Gasic, G. P., Kennedy, D. N., Hodge, S. M., Kaiser, J. R., Lee, M. J., Kim, B. W., Blood, A. J., Evins, A. E., Seidman, L. J., Iosifescu, D. v., Lee, S., Baxter, C., Perlis, R. H., Smoller, J. W., Fava, M., & Breiter, H. C. (2008). Cortical thickness abnormalities in cocaine addiction – a reflection of both drug use and a pre-existing disposition to drug abuse? *Neuron*, *60*(1), 174. https://doi.org/10.1016/J.NEURON.2008.08.011

Markowitz, H. (1952). Portfolio selection. *The Journal of Finance*, *7*(1), 77–91. https://doi.org/10.1111/j.1540-6261.1952.tb01525.x

Mason, W., & Suri, S. (2012). Conducting behavioral research on Amazon's Mechanical Turk. *Behavior Research Methods*, *44*(1), 1–23. https://doi.org/10.3758/S13428-011-0124-6

Paolacci, G., Chandler, J., & Ipeirotis, P. G. (2010). Running experiments on Amazon Mechanical Turk. *Judgement and Decision Making*, *5*(5), 411–419.

Perlis, R. H., Holt, D. J., Smoller, J. W., Blood, A. J., Lee, S., Kim, B. W., Lee, M. J., Sun, M., Makris, N., Kennedy, D. K., Rooney, K., Dougherty, D. D., Hoge, R., Rosenbaum, J. F., Fava, M., Gusella, J., Gasic, G. P., & Breiter, H. C. (2008). Association of a Polymorphism Near CREB1 With Differential Aversion Processing in the Insula of Healthy Participants. *Archives of General Psychiatry*, *65*(8), 882–892. https://doi.org/10.1001/ARCHGENPSYCHIATRY.2008.3

Robinson, T. E., & Berridge, K. C. (1993). The neural basis of drug craving: An incentive-sensitization theory of addiction. *Brain Research Reviews*, *18*(3), 247–291. https://doi.org/10.1016/0165-0173(93)90013-P

Schneirla, T. C. (1959). An evolutionary and developmental theory of biphasic processes underlying approach and withdrawal. In *Nebraska symposium on motivation, 1959.* (pp. 1–42). Univer. Nebraska Press.

Schneirla, T. C. (1965). Aspects of Stimulation and Organization in Approach/Withdrawal Processes Underlying Vertebrate Behavioral Development. *Advances in the Study of Behavior*, *1*(C), 1–74. https://doi.org/10.1016/S0065-3454(08)60055-8

Shannon, C. E., & Weaver, W. (1949). *The Mathematical Theory of Communication* (Vol. 1). University of Illinois Press.

Spielberger, C. D., Gorsuch, L., Laux, L., Glanzmann, P., & Schaffner, P. (2001). Das state-trait-angstinventar: STAI. *Beltz Test*.

Strauss, M. M., Makris, N., Aharon, I., Vangel, M. G., Goodman, J., Kennedy, D. N., Gasic, G. P., & Breiter, H. C. (2005). fMRI of sensitization to angry faces. *NeuroImage*, *26*(2), 389–413. https://doi.org/10.1016/J.NEUROIMAGE.2005.01.053

Tom, S. M., Fox, C. R., Trepel, C., & Poldrack, R. A. (2007). The Neural Basis of Loss Aversion in Decision-Making Under Risk. *Science*, *315*(5811), 515–518. https://doi.org/10.1126/science.1134239

Tversky, A., & Kahneman, D. (1992). Advances in prospect theory: Cumulative representation of uncertainty. *Journal of Risk and Uncertainty*, *5*(4), 297–323. https://doi.org/10.1007/BF00122574

Viswanathan, V., Lee, S., Gilman, J. M., Kim, B. W., Lee, N., Chamberlain, L., Livengood, S. L., Raman, K., Lee, M. J., Kuster, J., Stern, D. B., Calder, B., Mulhern, F. J., Blood, A. J., & Breiter, H. C. (2015). Age-related striatal BOLD changes without changes in behavioral loss aversion. *Frontiers in Human Neuroscience*, *9*(APR), 1–12. https://doi.org/10.3389/FNHUM.2015.00176/BIBTEX

Viswanathan, V., Sheppard, J. P., Kim, B. W., Plantz, C. L., Ying, H., Lee, M. J., Raman, K., Mulhern, F. J., Block, M. P., Calder, B., Lee, S., Mortensen, D. T., Blood, A. J., & Breiter, H. C. (2017). A quantitative relationship between signal detection in attention and approach/avoidance behavior. *Frontiers in Psychology*, *8*(FEB), 122. https://doi.org/10.3389/FPSYG.2017.00122/BIBTEX

Yamamoto, R., Ariely, D., Chi, W., Langleben, D. D., & Elman, I. (2009). Gender Differences in the Motivational Processing of Babies Are Determined by Their Facial Attractiveness. *PLOS ONE*, *4*(6), e6042. https://doi.org/10.1371/JOURNAL.PONE.0006042

Zhang, R., Brennan, T. J., & Lo, A. W. (2014). The origin of risk aversion. *Proceedings of the National Academy of Sciences*, *111*(50), 17777–17782. https://doi.org/10.1073/pnas.1406755111